\documentclass[aps, prd, amsfonts, pt12,floatfix]{revtex4}

\usepackage{graphicx}
\usepackage{amsmath,amsfonts}
\usepackage{epsfig,color}

\usepackage{hyperref} 

\renewcommand{\imath}{i}
\newcommand{\diag}{{\rm diag\,}}
\newcommand{\tr }{{\rm tr\,}}

\newcommand{\Ort}{{\rm O\,}}

\newcommand{\SO}{{\rm SO\,}}
\newcommand{\USp}{{\rm USp\,}}
\newcommand{\U}{{\rm U\,}}
\newcommand{\SU}{{\rm SU\,}}
\newcommand{\Herm}{{\rm Herm\,}}

\newcommand{\RE}{{\rm Re\,}}

\newcommand{\IM}{{\rm Im\,}}
\newcommand{\del}{\partial}
\newcommand{\eins}{\leavevmode\hbox{\small1\kern-3.8pt\normalsize1}}
\newcommand{\be}{\begin{eqnarray}}
\newcommand{\ee}{\end{eqnarray}}
\newcommand{\bmat}{\left ( \begin{array}{cc} }
\newcommand{\emat}{\end{array} \right ) }
\newcommand{\bvec}{\left ( \begin{array}{c} }
\newcommand{\evec}{\end{array} \right ) }
\newcommand{\nn}{\nonumber}

\begin{document}

\title{Dirac Spectra of  2-dimensional QCD-like theories}
\author{Mario Kieburg$^{1,2}$}
\email{mkieburg@physik.uni-bielefeld.de}
\author{Jacobus J. M. Verbaarschot$^{1}$}
\email{jacobus.verbaarschot@stonybrook.edu}
\author{Savvas Zafeiropoulos$^{1,3}$}
\email{zafeiropoulos@clermont.in2p3.fr}
\affiliation{$^{1}$ Department of Physics and Astronomy, Stony Brook University, Stony Brook, NY 11794-3800, USA}
\affiliation{$^{2}$ Fakult\"at f\"ur Physik, Universit\"at Bielefeld, Postfach 100131, 33501 Bielefeld, Germany}
\affiliation{$^{3}$ Laboratoire de Physique Corpusculaire, Universit\'e Blaise Pascal, CNRS/IN2P3, 63177 Aubi\`ere Cedex, France}

\date{\today}
\begin{abstract}
We analyze Dirac spectra of two-dimensional QCD like theories both in the 
continuum and on the lattice and classify them according to random matrix
theories sharing the same global symmetries. The classification is different
from QCD in four dimensions because the anti-unitary symmetries do not
commute with $\gamma_5$. Therefore in a chiral basis, the number of degrees of
freedom per matrix element are not given by the Dyson index. Our predictions
are confirmed by Dirac spectra from 
quenched lattice simulations for QCD with  two or three 
colors with
quarks in the fundamental representation as well as in the adjoint representation. The universality class of the spectra 
depends on the parity of the number of lattice points in each direction.
Our results show an agreement with random matrix theory that is qualitatively
similar to the agreement found for QCD in four dimensions. We discuss the implications
for the Mermin-Wagner-Coleman theorem and put our results in the context of
two-dimensional disordered systems.

\end{abstract}
\maketitle
\section{Introduction}

It has been well established that chiral symmetry is spontaneously broken
in strongly interacting systems of quarks and gluons for a wide range of
parameters such as the temperature, the chemical potential, the number
of colors, the number of flavors, the representation of the gauge group.
In the broken phase the corresponding low energy effective theory is given
by a weakly interacting system of pseudo-Goldstone bosons with a Lagrangian
that is determined by the pattern of chiral symmetry breaking. 
In lattice QCD the spontaneous breaking of chiral symmetry is studied
by evaluating the Euclidean partition function which is the 
average of the determinant of the Euclidean Dirac operator weighted by the Euclidean 
Yang-Mills action.  Its low energy limit
is given by the partition function of the Euclidean chiral Lagrangian. 
This theory simplifies drastically \cite{GL,LS} in the limit that the pion Compton wave-length is
much larger than the size of the box. Then the
 partition function factorizes into a part comprising the modes with zero momentum and a part describing the modes with non-zero
momentum. It turns out that the zero momentum part is equivalent to a random matrix theory with the same global symmetries of QCD \cite{VPLB}.

A particular useful way to study
chiral symmetry breaking is to analyze the properties of
 the eigenvalues of the Dirac operator. Because of the Banks-Casher
formula \cite{BC} the chiral condensate $\Sigma=|\langle \bar \psi \psi\rangle|$, the order parameter for the spontaneous breaking of chiral symmetry, 
is given by the average spectral density (denoted by $\rho(\lambda)$) near zero of the Dirac operator per unit of
the space-time volume $V$, 
\be\label{chiralcondensate}
\Sigma \equiv |\langle \bar \psi \psi\rangle| =
\lim_{a\to 0} \lim_{m\to 0} \lim_{V\to \infty}  
\frac 1V  \int_{-\infty}^\infty 
\frac {2m \rho(\lambda) d\lambda }{\lambda^2+m^2}  
= \lim_{a\to 0} \lim_{\lambda\to 0} \lim_{V\to \infty}   \frac \pi V \rho(\lambda).
\ee
Here,  $a$ is the lattice spacing which provides
the ultraviolet cut-off.
The order of the limits is critical and a different order gives a different
result. A better understanding of these limits can be obtained from the
behavior of the eigenvalue density of the Dirac operator on the scale
of the smallest eigenvalues which according to the Banks-Casher 
formula is given by
\be
\Delta \lambda = \frac 1{\rho(0)} = \frac \pi {\Sigma V}.
\ee
The so called microscopic spectral density is defined by \cite{SV}
\be
\rho_s(x) = \lim_{V\to \infty} \frac 1{\Sigma V}\rho\left ( \frac x{\Sigma V} \right ).   
\ee

If the Compton wavelength associated with the  
Dirac eigenvalues, $\lambda$, is much larger than
the size of the box, $L$, then
the partition  function that generates the  Dirac spectrum  factorizes
into a zero momentum part and a nonzero momentum part. The zero momentum
part is completely determined by the global symmetries of the QCD-(like) 
partition function and is equivalent to a random matrix theory with
the same global symmetries. 
The Compton wave length associated with the Dirac eigenvalues is the
Compton wave-length of the corresponding pseudo-Goldstone modes and
is given by
\be
\frac {2\pi}{m_\pi} = \frac {2\pi F_\pi}{\sqrt{2\lambda \Sigma}},
\ee 
where $F_\pi$ is the pion decay constant. The condition $1/m_\pi \gg L$ implies
\be
\lambda \ll \frac {F_\pi^2}{2 L^2 \Sigma}=\lambda_L,
\ee 
meaning that $\lambda_L$ is the characteristic eigenvalue scale 
 corresponding to the size of the box.  In $d$ dimensions the Euclidean volume is $V=L^d$ so that the average number of
eigenvalues in the universal domain scales as
\be
E_{\rm Th}\propto\frac{\lambda_L}{\Delta\lambda}=2\pi{F_\pi^2} L^{d-2}.
\ee
This scale is also known as the Thouless energy $E_{\rm Th}$. In two dimensions, the number
of eigenvalues in the universal domain remains of $O(1)$ 
in the thermodynamic limit.

Arguments have been made that in one dimensional systems all states become localized for an arbitrary small amount
of disorder. The two dimensional case is marginal. For site disorder 
all states are exponentially localized whereas for link  disorder 
the situation is less clear \cite{markos}. For systems that are both rotational invariant and time reversal invariant (denoted by the Dyson index $\beta_{\rm D}=1$) all states seem to be localized. In the case of
rotational invariant systems with broken time reversal invariance  
(denoted by the Dyson index $\beta_{\rm D} =2$)
states in the center of the band seem to be delocalized and the localization
length may be very large in a region around the band center.
For non-rotational invariant spin 1/2 systems 
(denoted by the Dyson index $\beta_{\rm D} =4$)
states are delocalized
for a substantial range of disorder and energies \cite{evangelou,furusaki,asada}.

The connection between localization and Goldstone bosons was most clearly
formulated by McKane and Stone \cite{stone}. They argued
 that a nonzero density of states around the origin for
a disordered system may either indicate the presence of Goldstone
bosons or may be due to a nonzero  density of the localized states. For dynamical
quarks the second alternative is not possible. The reason is that 
the eigenvalues of localized
states are uncorrelated so that the partition function
\be
Z(m)=\left\langle \prod_k(i\lambda_k+m)\right\rangle
\ee
factorizes into single eigenvalue partition functions
resulting in a vanishing 
chiral condensate \cite{jv-1997}.

If we take the results from the condensed matter literature at face value,
also in two dimensional systems,
there may be a finite region of extended states around zero with correlations
that are described by chiral random matrix theory or alternatively a partition function
with spontaneously broken chiral symmetry. 
In more than two dimensions we expect that
these  correlations will remain
in the presence of a fermion determinant and the corresponding partition function
will be the zero momentum part of a chiral Lagrangian. 
In two dimensions, the presence of a fermion determinant may
push these states beyond the Thouless
 energy so that {\it all} states become
localized. 
 This would reconcile the numerical results
for  $\beta_{\rm D} =4$ with the
Mermin-Wagner-Coleman theorem which states that a continuous symmetry cannot be broken spontaneously
in two or less dimensions in systems with sufficiently short-range interactions. 
In terms of the supersymmetric formulation of the quenched limit,
the Mermin-Wagner-Coleman theorem  could be evaded because the 
symmetry group is non-compact. This has been shown for hyperbolic
spin models \cite{seiler,seiler2}.  
This opens the possibility to have extended states and  universal  
spectral correlations also in two dimensions.

Another interpretation is possible if the localization length $\xi$ is large 
so that we can consider the limiting case
\be
1 \ll L \ll \xi.
\label{spont}
\ee
Then the states behave as extended states with eigenvalues that 
are described by random matrix theory up to the Thouless energy. In 
this case the problems with the Mermin-Wagner-Coleman theorem can be avoided
and a  transition to a localized phase only takes place when $L \sim \xi$.
In such scenario the scalar correlation  function may drop off at a similar rate, so
that chiral symmetry appears to be broken in the regime  \eqref{spont}.
This correlation function was studied for the $N_{\rm c}$-color Thirring model
\cite{witten-thir} with a drop-off of $1/x^{1/N_{\rm c}}$.

A two-dimensional model for which Dirac spectra
have  been studied in great detail,
both analytically and numerically, is the Schwinger model. The eigenvalue
correlations of the  one-flavor Schwinger model are given by random matrix theory
as was shown numerically \cite{farchioni,damgaard} and analytically by
calculating the Leutwyler-Smilga sum rules \cite{leonid}. The two flavor
Schwinger model was analyzed in great detail in \cite{bietenholtz}, 
and after rescaling the eigenvalues by the average level spacing excellent
agreement with chiral random matrix theory is observed.  
No agreement with chiral random matrix theory is found for the quenched
Schwinger model \cite{damgaard}, while the spectral density seems 
to diverge for
$\lambda \to 0$.  The repulsion between the eigenvalues seems to be
greatly suppressed indicating that the states are localized.

Let us consider a theory where the mass dependent chiral condensate scales with
the quark mass as
\be
\hspace*{-2cm}\Sigma(m) \sim m^\alpha. 
\ee
In the Schwinger model we have that \cite{smilga1}
\be\label{alpha}
\alpha= \frac{N_{\rm f}-1}{N_{\rm f}+1},
\ee
but the argument given in this paragraph is more general.
According to the Gell-Mann-Oakes-Rennner relation the mass of the "pions''
associated with this condensate is given by
\be\label{mpi}
m_\pi^2 \sim m^{\alpha+1}.
\ee
Using the relation~\eqref{chiralcondensate} between the spectral density and the chiral condensate
we  find that the eigenvalue density behaves as
\be
\rho(\lambda) \sim V\Sigma(\lambda)\sim V \lambda^\alpha.
\ee
The Thouless energy is given by the scale for which the pion Compton wavelength
is equal to the size of the box, i.e. $m_\pi\propto1/L$. Employing the relation~\eqref{mpi} we find the mass associated to the Thouless energy,
\be\label{thmass}
m_{\rm th} \sim L^{-2/(\alpha+1)}.
\ee
The integrated spectral density is given by
\be\label{thnumber}
N(\lambda)=\int_{-\lambda}^\lambda\rho(\lambda')d\lambda' \sim V \lambda^{\alpha+1},  
\ee
so that the average number of eigenvalues below the Thouless energy is proportional to
\be
N_{\rm th}=N(m_{\rm th} ) \sim L^{d-2}.
\ee
due to combination of Eqs.~\eqref{thmass} and \eqref{thnumber}. Remarkably, the number of eigenvalues described by random matrix theory
does not depend on $\alpha$. In two dimensions this number is constant
in the thermodynamic limit
but the agreement with chiral
random matrix theory seems to improve with larger volumes for the Schwinger model \cite{damgaard}. 
The corollary of this argument is that correlations
of low-lying Dirac eigenvalues in conformal QCD-like theories are given
by chiral random matrix theory after unfolding the eigenvalues, i.e.
$\lambda_k' = \lambda_k^{\alpha +1}$ .

The eigenvalues scale with the volume as
\be
\lambda \sim V^{-1/(\alpha+1)}
\ee
via the relation~\eqref{thnumber} when keeping the average number of eigenvalues fixed. This scaling was studied in \cite{bietenholtz} where a  volume 
scaling of $V^{-5/8}$ is observed for two almost massless flavors, c.f. Eq.~\eqref{alpha}. 
This would correspond to $N_{\rm f} =4$, cf. Eq.~\eqref{alpha}. This is actually correct because the lattice Dirac operator couples only even and odd sites doubling the number
of flavors. Apparently, we need exact massless quarks to push the
states in the localized domain.

Another important difference between QCD in four dimensions and QCD in lower dimensions
is the index of the Dirac operator. In four dimensions the index is
equal to the topological charge of the gauge field configurations. In three dimensions the index
is not defined. In two dimensions topology is defined for $\U(1)$ and can for example
be studied for the Schwinger model \cite{smil2,leonid}. However, for higher dimensional
gauge groups the index of the Dirac operator is zero 
\cite{smilga1,Christiansen:1998ax,Christiansen:1997mu}
although unstable instantons do exist \cite{witten2d,Gross:1994mr}.

In this paper we consider the quenched two-dimensional QCD Dirac 
operator in the strong coupling limit with the gauge fields distributed according to the 
Haar measure. Both the continuum limit and the lattice QCD Dirac operator
will be discussed. For the lattice Dirac operator we employ naive fermions. 
Our original motivation 
for this choice was to understand the transition between different symmetry classes when taking the continuum limit which was observed for staggered fermions in three \cite{Burda} as well as in four \cite{BKPW} dimensions, but this
issue is not addressed in this paper.

The strong coupling lattice model is expected to be equivalent to 
an interacting theory of mesons and/or baryons. For U(1) gauge theories in
two dimensions this has been shown explicitly \cite{simons-altland}
by means of a color-flavor transformation \cite{zirn-cf,wettig-wei,wettig-sl1}, 
where a gradient
expansion generates the various terms of a chiral Lagrangian. 
In this paper, we do not perform the continuum limit, so that the lattice
theory is equivalent to an unrenormalized chiral Lagragian, and 
 the usual arguments, that the states below the Thouless energy  
are correlated according to
random matrix theory, apply. In the continuum limit and two dimensions, 
the fluctuations of the hadronic fields will
dominate the chiral condensate and the theory renormalizes to a trivial phase
without Goldstone bosons. In other words, the theory renormalizes to a 
localized phase.

The symmetry breaking pattern for the continuum limit in any dimension
was discussed in \cite{imbo} and in the context of topological insulators in \cite{ludwig}
and goes back to what is known as Bott-periodicity. The dimensional dependence
of the symmetry breaking pattern (see Table \ref{table1}) has its origins in the
structure of  Clifford algebras. The four dimensional symmetry breaking pattern and
its description in terms of random matrix theory has been known
for a long time \cite{shifman-3,peskin,V}. Because of the absence of the $\gamma_5$ Dirac matrix in
three dimensions the symmetry breaking pattern is different \cite{VZ3,magnea1,magnea4}. 
In two dimensions, the  $\gamma_5$ matrix is replaced by the third Pauli matrix, $\sigma_3$. This matrix  also anticommutes with the Dirac operator,
but it does not commute with the charge conjugation matrix given by $\sigma_2$,
which leads 
to a different symmetry breaking pattern \cite{imbo}.

The symmetries of the Dirac operator also depend on the parity of the lattice.
If the lattice size is even in both directions, even lattice sites are only
coupled to odd lattice sites resulting in a ``lattice chiral symmetry''. Having
a lattice that is odd in one direction and even in the other one also
puts global constraints on the Dirac operator resulting in different symmetry
properties. In this paper we classify lattice theories in terms of random
matrix theory. In total we can distinguish 9 classes. First of all they differ in their anti-unitary symmetries, namely with no
anti-unitary symmetry, with an anti-unitary symmetry that squares to 1, and
with an anti-unitary symmetry that squares to -1. Moreover for each of these three
classes we can have an even-even, an even-odd or an odd-odd lattice.
For all nine classes we give the spectral properties in the microscopic 
domain and compare them with lattice simulations of the corresponding
lattice theory in the strong coupling limit.

In Section 2 we discuss the microscopic Dirac spectrum and chiral symmetry breaking
pattern for the continuum limit of two-dimensional QCD. The two dimensional lattice
gauge theory for three different values of the Dyson index is analyzed in section 3,
and concluding remarks are made in Section 4. In the appendices we derive several
random matrix results that have been used in the main text.

\section{Continuum Dirac Operator}\label{sec:cont}

The Euclidean Dirac operator of QCD-like theories is given by
\begin{eqnarray}\label{1.1}
 \mathcal{D}=\gamma^\mu(\partial_\mu+ \imath A_\mu^a {\lambda}_a),
\end{eqnarray}
where $A_\mu^a$ are the gauge fields, $\gamma_\mu$ are the Euclidean 
$\gamma$-matrices and $\lambda_a$ are the generators of the gauge group. For an even number of dimensions the Dirac operator in a 
 chiral basis  reduces to a  $2\times2$ block structure 
\begin{eqnarray}\label{1.4}
 \mathcal{D}^{(2/4)} =\left[\begin{array}{cc} 0 & \mathcal{W}^{(2/4)} \\ -\mathcal{W}^{(2/4)\,\dagger} & 0 \end{array}\right],
\end{eqnarray}
where in two dimensions the operator $\mathcal{W}^{(2)}$ is given by
\begin{eqnarray}\label{1.5}
 \mathcal{W}^{(2)}=\partial_1+\imath\partial_2+(\imath A_1^a-A_2^a){\lambda}_a,
\end{eqnarray}
 and in four dimensions the operator $\mathcal{W}^{(4)}$ can be written as
\begin{eqnarray}\label{1.6}
 \mathcal{W}^{(4)}&=&\imath\sigma_\mu(\del_\mu+\imath A_\mu^a \lambda_a)
\end{eqnarray}
employing the standard chiral representation of Euclidean 
$\gamma$-matrices with $\sigma_\mu = (\sigma_k,-\imath\eins_4)$ and $\sigma_k$ 
the Pauli matrices.

In three dimensions, the Dirac operator is given by
\be
 \mathcal{D}=\sum_{k=1}^3\sigma^k(\partial_k+ \imath A_k^a {\lambda}_a).
 \ee
There is no involution that anti-commutes with the Dirac operator so that there is
 no chiral block structure.
This is the crucial difference with even dimensional theories and was already studied in Ref.~\cite{VZ}.

In addition to chiral symmetry the Dirac operator  has  other symmetries depending on the representation of the gauge group which is discussed in the ensuing subsections. 
In subsection~\ref{sec:cont.1} we recall the discussion of the global symmetries of the QCD Dirac operator in three and four dimensions and extend it to the two dimensional theory as well. This symmetry classification is summarized in table~\ref{table1}. In subsection~\ref{sec:cont.2} we discuss the corresponding random matrix theories. Thereby we summarize the classification of the random matrix theories for three- and four-dimensional continuum QCD and  supplement this  with the random matrix theories for two-dimensional QCD. In subsection~\ref{sec:cont.3}
we recall the symmetry breaking patterns.

\begin{table}
\begin{tabular}{|c|c|c|c|c|c|}\hline
\quad d\quad\ & \quad Gauge theory\quad\ & \quad $\beta_{\rm D} $ \quad\  & \quad $\alpha_{\rm D}$ \quad\ & Symmetry Breaking Pattern & RMT \\
\hline\hline
2   & $N_{\rm c}=2$, fund. & $1$  & $1$ &   \quad$\USp(2N_{\rm f}) \times \USp(2N_{\rm f}) \to \USp(2N_{\rm f})$\quad &  (CI) \\  
2   & $N_{\rm c}\ge 3$, fund.  & $2$  & $1$ &   $\U(N_{\rm f})\times \U(N_{\rm f}) \to \U(N_{\rm f})$& \ chGUE (AIII)\ \\
2 & $N_{\rm c}\geq2$, adj.   & $4$  & $1,5$ &   $\Ort(2N_{\rm f}) \times \Ort(2N_{\rm f}) \to \Ort(2N_{\rm f})$&   (DIII) \\
\hline
3   & $N_{\rm c}=2$, fund.   & $1$ & $0$   &$\USp(4N_{\rm f}) \to \USp(2N_{\rm f}) \times \USp(2N_{\rm f})$ & GOE   (AI)\\ 
3 & $N_{\rm c}\ge 3$, fund.  & $2$ & $0$   &$\U(2N_{\rm f}) \to \U(N_{\rm f}) \times \U(N_{\rm f})$   &GUE (A)\\
3 & $N_{\rm c}\geq2$, adj.    & $4$ & $0$   & $\Ort(2N_{\rm f}) \to \Ort(N_{\rm f}) \times \Ort(N_{\rm f})$& GSE (AII)\\
\hline
4   & $N_{\rm c}=2$, fund.   & $1$ & $\nu$ &   $\U(2N_{\rm f})\to\USp(2N_{\rm f})$ & chGOE (BDI)\\
4 & $N_{\rm c}\ge 3$, fund.  & $2$ & $2\nu+1$ &   $\U(N_{\rm f})\times \U(N_{\rm f}) \to \U(N_{\rm f})$ & chGUE (AIII)\\
4 & $N_{\rm c}\geq2$, adj.    &  $4$ & $4\nu+3$ &  $\U(2N_{\rm f})\to \Ort(2N_{\rm f})$ & chGSE  (CII)\\ \hline
\end{tabular}
 \caption[]{Symmetry breaking patterns in two ($d=2$), three ($d=3$), and four ($d=4$) dimensions for 
different gauge theories and their associated Dyson index $\beta_{\rm D}$ which is equal to the level repulsion.
The corresponding random matrix
theory sharing the same symmetry breaking pattern and its classificatrion according to symmetric spaces is indicated in the last column. 
The repulsion of the levels from the origin, $\lambda^{\alpha_{\rm D}}$, depends on the topological charge $\nu$ for  QCD-like theories in four dimensions.  
The case of the two-dimensional $\SU(N_{\rm c})$ theory with the fermions in the adjoint representation is particular since the index of the Dirac operator is
either $0$ or $1$ depending on the parity of the dimensions of the Dirac matrix. This
results in a repulsion that is either $\lambda$ or $\lambda^5$. 
The corresponding random matrix theory consists of anti-symmetric
off-diagonal blocks so that depending on the dimensionality we have 
either  no or one pair of generic zero modes, respectively. For a discussion of the classification of random matrix theories in terms of symmetric spaces we refer to Refs.~\cite{dyson-3,class}. In this table we do not include the
breaking of the axial symmetry. 
\label{table1}}
\end{table}

\subsection{Anti-unitary symmetries of the QCD Dirac operator} \label{sec:cont.1}

The anti-unitary symmetries of the Dirac operator depend on the representation of the generators $\lambda_a$ of the gauge group $\SU(N_{\rm c})$. We consider three different gauge theories, namely with the gauge group $\SU(N_{\rm c}=2)$ and
 fermions in  the fundamental representation denoted by the Dyson index $\beta_{\rm D}=1$,  with the gauge group $\SU(N_{\rm c}>2)$ and fermions in  the fundamental representation which is $\beta_{\rm D}=2$, and with the gauge group $\SU(N_{\rm c}\geq2)$ and  fermions in  the adjoint representation labelled by $\beta_{\rm D}=4$.

\subsubsection{$\beta_{\rm D}=1$} \label{sec:cont.1a}

Let us consider the first case which is QCD with two colors ($N_{\rm c}=2$) and
fermions in  the fundamental representation. Then the $\lambda_a$  are given by the three Pauli matrices $\tau_a$ acting in color space.
Hence each covariant derivative
\begin{eqnarray}\label{1.7}
 \mathcal{D}_\mu=\partial_\mu+\imath A_\mu^a {\tau}_a
\end{eqnarray}
is pseudo-real (quaternion) and anti-Hermitian, i.e.
\begin{eqnarray}\label{1.8}
 \mathcal{D}_\mu^\dagger=-\mathcal{D}_\mu\quad {\rm and}\quad  [\mathcal{D}_\mu,\tau_2K]_-=\mathcal{D}_\mu\tau_2K-\tau_2K\mathcal{D}_\mu=0
\end{eqnarray}
with $K$ the complex conjugation operator. The corresponding Dirac operator has the anti-unitary symmetry
\be\label{sym-bet1}
[i\mathcal{D}^{(d)}, \tau_2 C K]_-=0, 
\ee
where $C$ is the charge conjugation matrix. In four dimensions the charge conjugation matrix reads
$C\equiv \gamma_2\gamma_4$ and in two and three dimensions it is 
given by
$C=\sigma_2$.  

A crucial point is that the anti-unitary operator satisfies
\be\label{anti-uni1}
( C\tau_2 K)^2=1.
\ee
Therefore one can always find a gauge field independent basis for which the Dirac operator is real \cite{dyson-3,V}. 
This is the reason why this case is denoted by the Dyson index $\beta_{\rm D} =1$ (one degree of freedom per matrix element).
Collecting everything, the continuum 
Euclidean QCD Dirac operator for QCD with two fundamental fermions   fulfills three global symmetries in four and two dimensions namely anti-Hermiticity, chiral symmetry, and a reality condition, i.e.
\begin{eqnarray}\label{symm-bet1-4d}
 \mathcal{D}^{(4)\,\dagger}=-\mathcal{D}^{(4)},\quad [\mathcal{D}^{(4)},\gamma_5]_+=0,\quad {\rm and}\quad  [i\mathcal{D}^{(4)},\tau_2\gamma_2\gamma_4K]_-=0
\end{eqnarray}
for four dimensions, see Ref.~\cite{V}, and
\begin{eqnarray}\label{symm-bet1-2d}
 \mathcal{D}^{(2)\,\dagger}=-\mathcal{D}^{(2)},\quad [\mathcal{D}^{(2)},\sigma_3]_+=0,\quad
 {\rm and}\quad  [i\mathcal{D}^{(2)},\tau_2 \sigma_2K]_-=0
\end{eqnarray}
for two dimensions. For three dimensions there is no chiral symmetry but the rest remains the same as in the even dimensional case
\begin{eqnarray}\label{symm-bet1-3d}
 \mathcal{D}^{(3)\,\dagger}=-\mathcal{D}^{(3)}\quad {\rm and}\quad  [\imath\mathcal{D}^{(3)},\tau_2\sigma_2K]_-=0,
\end{eqnarray}
see Ref.~\cite{VZ}.
 Next we  discuss the implications of these symmetries.

In four dimensions, Eq.~\eqref{symm-bet1-4d} implies that 
we can construct a gauge field independent basis for which 
the Dirac operator decomposes into a chiral block structure or a basis
for which the Dirac operator becomes real. This can be done at the same time
if the 
projection onto a chiral basis commutes with the anti-unitary symmetry.
This is the case in four dimensions where
\be\label{comm-bet1-4d}
\left[\frac{1\pm\gamma_5}2,\tau_2\gamma_2\gamma_4K\right]_-=0.
\ee
 The corresponding random matrix ensemble is the chiral Gaussian orthogonal ensemble (chGOE), see Refs.~\cite{V}.

Equation~\eqref{comm-bet1-4d} does not carry over to the two-dimensional theory. In this case the projectors onto a chiral basis are given by  
$(1\pm\sigma_3)/2$, playing the role of $(1\pm\gamma_5)/2$, but the
commutator with the anti-unitary operator does not vanish,
\be\label{comm-bet1-2d}
\left[\frac{1\pm \sigma_3}2,\tau_2\sigma_2K\right]_-\neq 0.
\ee
Therefore, one cannot find a basis for which the two-dimensional Dirac operator decomposes into
real chiral blocks.

 Choosing the chiral basis for $\mathcal{D}^{(2)}$ the anti-unitary symmetry 
yields a different condition
\be\label{symmetry-cont-bet1a}
\left [ \bmat 0 & \imath \tau_2K \\-\imath \tau_2K & 0 \emat  ,
\bmat 0 & \imath\mathcal{W}^{(2)} \\ -\imath\mathcal{W}^{(2)\,\dagger} & 0 \emat \right ]_-=0,
\ee  
which is equivalent to
\be\label{symmetry-cont-bet1b}
\mathcal{W}^{(2)}= -\tau_2 \mathcal{W}^{(2)\,T} \tau_2.
\ee
Thus the operator is anti-self-dual and complex since we have no additional symmetries. After a unitary transformation one obtains an equivalent Dirac operator with an off-diagonal block $\tau_2 \mathcal{W}^{(2)}$ which is complex symmetric. The corresponding random matrix is known as the first Bogolyubov-de Gennes ensemble denoted by the Cartan symbol CI, see Ref.~\cite{class}, and has been applied to the normal-superconducting transitions in mesoscopic physics~\cite{az}.

In three dimensions  we can construct a gauge field
independent basis for which the matrix elements of the 
operator $\imath\mathcal{D}^{(3)}$ become real symmetric. The corresponding random matrix ensemble is  the Gaussian orthogonal ensemble (GOE), see Refs.~\cite{VZ}.

\subsubsection{$\beta_{\rm D}=2$} \label{sec:cont.1b}

In the case of three or more colors ($N_{\rm c}\geq3$) with the fermions in the fundamental 
representation the symmetry under  complex conjugation~\eqref{1.8} is lost. 
Only anti-Hermiticity and, for even dimensions, chiral symmetry 
survive.  The global symmetries of the Dirac operator are
\begin{eqnarray}\label{symm-bet2-2d}
 \mathcal{D}^{(2)\,\dagger}=-\mathcal{D}^{(2)}\quad{\rm and}\quad[\mathcal{D}^{(2)},\sigma_3]_+=0
\end{eqnarray}
in two dimensions and
\begin{eqnarray}\label{symm-bet2-4d}
 \mathcal{D}^{(4)\,\dagger}=-\mathcal{D}^{(4)}\quad{\rm and}\quad[\mathcal{D}^{(4)},\gamma_5]_+=0
\end{eqnarray}
in four dimensions. Since there are no anti-unitary symmetries
 the operator $\mathcal{W}^{(2/4)}$ is generically complex both in two and
four dimensions. 
This is the reason why we denote this case by the Dyson index
 $\beta_{\rm D} =2$.
Therefore  the random matrix ensemble corresponding to
the Dirac operator $\mathcal{D}^{(2)}$ as well as $\mathcal{D}^{(4)}$ 
is given by an ensemble of chiral, complex, anti-Hermitian random matrices which can be chosen with 
Gaussian weights. This ensemble is known as the chiral Gaussian Unitary Ensemble (chGUE), see Refs.~\cite{V}.

In three dimensions we only have the anti-Hermiticity condition,
\begin{eqnarray}\label{symm-bet2-3d}
 \mathcal{D}^{(3)\,\dagger}=-\mathcal{D}^{(3)}.
\end{eqnarray}
Hence the operator $\imath\mathcal{D}^{(3)}$ is Hermitian and its analogue in random matrix theory is the Gaussian Unitary Ensemble (GUE). 
The three dimensional case was discussed 
in Ref.~\cite{VZ} and its predictions for the microscopic Dirac spectrum
have been confirmed by various lattice simulations~\cite{Burda}.

\subsubsection{$\beta_{\rm D}=4$} \label{sec:cont.1c}

The third case is for fermions in the adjoint representation with two or more colors ($N_{\rm c}\geq2$). In this case the generators of the gauge group
are anti-symmetric and purely imaginary. This results in two symmetry relations for the covariant derivatives
\begin{eqnarray}\label{1.9}
 \mathcal{D}_\mu^\dagger=-\mathcal{D}_\mu\quad {\rm and}\quad [K,\imath\mathcal{D}_\mu]_-=0.
\end{eqnarray}
The corresponding Dirac operator fulfills the anti-unitary symmetry
\be\label{sym-bet4}
[\imath \mathcal{D}^{(d)},CK]_-=0,
\ee
where the anti-unitary operator satisfies
\be\label{anti-uni2}
( CK)^2=-1
\ee
for all dimensions. This allows us to construct
 a gauge field independent basis for which the matrix elements of the
Dirac operator can be grouped into real
quaternions. This case is denoted by the Dyson index $\beta_{\rm D} =4$. 

Collecting all global symmetries of the Dirac operator we have
\begin{eqnarray}\label{symm-bet4-4d}
 \mathcal{D}^{(4)\,\dagger}=-\mathcal{D}^{(4)},\quad [\mathcal{D}^{(4)},\gamma_5]_+=0,\quad {\rm and}\quad  [\imath\mathcal{D}^{(4)},\gamma_2\gamma_4K]_-=0
\end{eqnarray}
for four dimensions,
\begin{eqnarray}\label{symm-bet4-2d}
 \mathcal{D}^{(2)\,\dagger}=-\mathcal{D}^{(2)},\quad [\mathcal{D}^{(2)},\sigma_3]_+=0,\quad {\rm and}\quad  [\imath\mathcal{D}^{(2)},\sigma_2K]_-=0
\end{eqnarray}
for two dimensions, and
\begin{eqnarray}\label{symm-bet4-3d}
 \mathcal{D}^{(3)\,\dagger}=-\mathcal{D}^{(3)}\quad {\rm and}\quad  [\imath\mathcal{D}^{(3)},\sigma_2K]_-=0
\end{eqnarray}
for three dimensions. The last case is  the simplest.
There is no chiral symmetry, but  we can construct a basis for which
the matrix elements of the Hermitian operator 
$\imath \mathcal{D}^{(3)}$ can be grouped into real quaternions. 
The associated random matrix ensemble is the Gaussian Symplectic Ensemble (GSE) pointed out for the first time in Ref.~\cite{VZ}.

In two and four dimensions we have again to consider 
 the commutator of the projection operators onto the eigenspaces of $\gamma_5$
and  the anti-unitary operator. As is the case for $\beta_{\rm D} =1$, the commuator
vanishes in four dimensions,
\be\label{comm-bet4-4d}
\left[\frac{\eins\pm\gamma_5}{2},\gamma_2\gamma_4K\right]_-=0.
\ee
Therefore we can  construct a basis for  which $\mathcal{D}^{(4)}$ decomposes
into chiral blocks with quaternion real elements.  Therefore, such Dirac operators are in the universality class of the chiral Gaussian Symplectic Ensemble 
(chGSE), see Refs.~\cite{V}.

In two dimensions the commutator of the anti-unitary symmetry and the chiral
projector does not vanish, i.e.
\be\label{comm-bet4-2d}
\left[\frac{\eins\pm\sigma_3}{2},\sigma_2K\right]_-\neq0.
\ee
Therefore, there is no  gauge field independent basis for which 
$\mathcal{D}^{(2)}$ decomposes into quaternion real chiral blocks.
Nevertheless, we can find a basis for which one of these properties
holds.
 In a chiral basis the anti-unitary symmetry~\eqref{sym-bet4} reads
\be\label{symmetry-cont-bet4a}
\left [ \bmat 0 & \imath K \\-\imath K & 0 \emat  ,
\bmat 0 & \imath\mathcal{W}^{(2)} \\ -\imath\mathcal{W}^{(2)\,\dagger} & 0 \emat \right ]_-=0
\ee  
and results into
\be\label{symmetry-cont-bet4b}
\mathcal{W}^{(2)\,T} = -\mathcal{W}^{(2)}.
\ee  
Thus the operator $\mathcal{W}^{(2)}$, see Eq.~\eqref{1.5},
 is complex anti-symmetric. In random matrix theory this symmetry class is
 known as the second Bogolyubov-de Gennes ensemble denoted by the Cartan symbol DIII~\cite{class}.
 This ensemble also plays an important role in mesoscopic physics~\cite{az}.

\subsection{Random Matrix Theory for continuum QCD} \label{sec:cont.2}

As was outlined in Refs.~\cite{V,SV} a 
random matrix theory for the Dirac operator is obtained by replacing its 
matrix elements by random numbers while maintaining the 
 global unitary and anti-unitary symmetries of the QCD(-like) theory. 
Within a wide class, the distribution of the eigenvalues on the scale of
the average level spacing does not depend on the probability distribution
of the matrix elements. This allows us to choose the probability distribution
to be Gaussian.
The random matrix  partition function is thus given by
\be\label{chrmt}
Z_{N_{\rm f}}^\nu =\int d[D] \exp\left[ -\frac{n\beta_{\rm D}}2 \tr  D^\dagger D\right] 
\prod_{k=1}^{N_{\rm f}}\det (D+m_k\eins).  
\ee
In even dimensions, in particular for $d=2,4$, the Dirac operator
has the chiral block structure
\be\label{dirac}
D = \bmat 0 & \imath W \\ \imath W^\dagger & 0 \emat ,
\ee
while in three dimensions the Dirac operator is still anti-Hermitian
but the block structure is absent.  
The mass matrix for the $N_{\rm f}$ quarks is given by $M=\diag(m_1,\ldots,m_{N_{\rm f}})$.
 The measure $d[D]$  is the product of all real independent differentials of the matrix elements of $D$.

In three dimensions, the random matrix ensemble is $n\times n$ dimensional for $\beta_{\rm D}=1,2$ and $2n\times2n$ dimensional for $\beta_{\rm D}=4$. 
The random matrix 
$\imath D$ is either real symmetric ($\beta_{\rm D}=1$), Hermitian 
($\beta_{\rm D}=2$), or Hermitian self-dual ($\beta_{\rm D}=4$). From the corresponding joint probability density of the eigenvalues \cite{Mehtabook},
\be\label{joint-nonchi}
 p_{d=3}(\Lambda)\prod\limits_{1\leq j\leq n}d\lambda_j\propto|\Delta_{n}(\Lambda)|^{\beta_{\rm D}}
\prod\limits_{1\leq j\leq n}\exp\left[-\frac {n\beta_{\rm D}}2\lambda_j^2\right] d\lambda_j,
\ee
one can already read off many important spectral properties of the QCD-Dirac operator $\mathcal{D}^{(3)}$ in the microscopic limit, cf. table~\ref{table1}. Recall the Vandermonde determinant
\be\label{Van}
 \Delta_{n}(\Lambda)=\prod\limits_{1\leq a<b\leq n}(\lambda_a-\lambda_b)=(-1)^{n(n-1)/2}\det\left[\lambda_a^{b-1}\right]_{1\leq a,b\leq n}.
\ee
Thus, in three dimensions the eigenvalues are not degenerate apart from the Kramers degeneracy of QCD with adjoint fermions. Moreover, the eigenvalues of  $\mathcal{D}^{(3)}$ repel each other like $|\lambda_a-\lambda_b|^{\beta_{\rm D}}$ and have no repulsion from the origin~\cite{VZ}.

In four dimensions, the operator $\mathcal{W}^{(4)}$ is replaced by an $ n\times (n+\nu)$ real ($\beta_{\rm D}=1$) or complex ($\beta_{\rm D}=2$) random matrix $W$ or a $2n\times2(n+\nu)$ quaternion matrix for $\beta_{\rm D}=4$. 
Then the Dirac operator has exactly $\nu$ and $2\nu$ zero modes for 
$\beta_{\rm D}=1,2$ and $\beta_{\rm D}=4$, respectively. Therefore, $\nu$ is identified as the index of the Dirac operator. Due to 
the axial symmetry the nonzero eigenvalues always come in pairs 
$\pm\imath\lambda$. Moreover, because of the quaternion structure, the eigenvalues of $\mathcal{D}^{(4)}$ 
as well as of the corresponding random matrix Dirac operator  are degenerate 
for QCD with adjoint fermions. The joint probability density of the eigenvalues of the random matrix $D$ reads \cite{V}
\be\label{joint-chi}
 p_\chi(\Lambda)\prod\limits_{1\leq j\leq 2n}d\lambda_j\propto|\Delta_{2n}(\Lambda^2)|^{\beta_{\rm D}}
\prod\limits_{1\leq j\leq 2n}\exp\left[-\frac{n\beta_{\rm D}}2\lambda_j^2\right]\lambda_j^{\alpha_{\rm D}} d\lambda_j,
\ee
cf. table~\ref{table1}. Again we can read off the behavior of the eigenvalues of $D$ which, in the microscopic limit,  
are shared with the behavior of the low-lying 
eigenvalues of the QCD Dirac operator. 
The eigenvalues again repel each other as  $|\lambda_a-\lambda_b|^{\beta_{\rm D}}$. The difference with 
the three dimensional case is the level repulsion from the origin $\lambda_a^{\alpha_{\rm D}}=\lambda_a^{\beta_{\rm D}(\nu+1)-1}$ which results from the generic zero modes and the chiral structure of the Dirac operator. The global symmetries of the four-dimensional QCD Dirac operator and their impact on the microscopic spectrum were discussed in Refs.~\cite{SV,V}.

In two dimensions, rather than choosing a basis for which the Dirac operator
becomes real or quaternion real for $\beta_{\rm D}=1$ and $\beta_{\rm D} = 4$, respectively,
we insist on a chiral basis that preserves the chiral block structure of the
Dirac operator. This results in a random matrix theory for which the matrix
$\tau_2W$ is complex symmetric 
 for $\beta_{\rm D} =1$, $\tau_2W=(\tau_2W)^T \in\mathbb{C}^{2n\times 2n}$  and complex anti-symmetric
 for $\beta_{\rm  D} =4$, $W=-W^T\in\mathbb{C}^{n\times n}$, cf. Eqs.~\eqref{symmetry-cont-bet1b} and \eqref{symmetry-cont-bet4b}, respectively. 
For QCD with three or more colors and the fermions in the fundamental representation ($\beta_{\rm D} = 2$), the two-dimensional
Dirac operator has the same symmetries as the four-dimensional theory resulting in the same random matrix theory.

Another important difference between two and four dimensions is the topology of the gauge field configurations. For
 QCD with fundamental fermions the homotopy class is $\Pi_1(\SU(2))= 0$. Hence,
no stable instanton solutions
exist \cite{smilga,Christiansen:1997mu} (unstable instanton solutions are still 
possible \cite{witten2da,witten2d,Gross:1994mr}). Also the index of the Dirac
operator is necessarily zero.
Suppose that the two-dimensional Dirac operator has an exact zero mode
\be\label{zero-mode.a}
\mathcal{D}^{(2)}\phi =0
\ee
with definite chirality
\be\label{chirality.a}
\sigma_3\phi =\pm \phi.
\ee
Then, because of the anti-unitary symmetry, we also have that
\be\label{zero-mode.b}
\mathcal{D}^{(2)}  \sigma_2\tau_2K \phi =0,
\ee
which generates another zero mode unless $ \sigma_2\tau_2K \phi$ and $\phi$ are linearly dependent.
This exactly happens in the four-dimensional theory. However in two dimensions
$\phi$ and $ \sigma_2\tau_2K \phi$ have opposite chiralities
\be\label{chirality.b}
\sigma_3\sigma_2\tau_2K \phi =-\sigma_2\tau_2K \sigma_3 \phi=\mp\sigma_2\tau_2K \phi
\ee
implying that they
have to be linearly independent states. We conclude that the index of the Dirac operator
is zero for two-dimensional QCD  in the fundamental representation and with two colors.

Although the index is trivial we still have a linear repulsion of the spectrum from the origin resulting from the chiral structure of $\mathcal{D}$.
 The joint probability density of the 
corresponding random matrix ensemble was first derived in the context
of mesoscopic phyiscs \cite{az} and is 
given by  Eq.~\eqref{joint-chi}.  
For completeness we give a derivation of this result in appendix~\ref{app1.a}. Since we have a linear repulsion from the origin we have $\alpha_{\rm D}=1$. The level repulsion is 
also linear, i.e. $\sim |\lambda_a-\lambda_b|$, and the eigenvalues show no generic degeneracy.

For quarks in the adjoint representation the gauge group is given by $\SU(N_{\rm c})/\mathbb{Z}_{N_{\rm c}}$ with the
homotopy group $\Pi_1(\SU(N_{\rm c})/\mathbb{Z}_{N_{\rm c}}) = \mathbb{Z}_{N_{\rm c}}$ \cite{smilga}.  If $\phi $ is 
a zero mode with positive chirality, then $\sigma_2 K \phi$ is a zero mode with negative
chirality. Therefore, the index of the Dirac operator is  zero. Using a
bosonization approach it can be shown that the chiral condensate
is nonzero for all $N_{\rm c}$ \cite{smilga-cond}, which is consistent with having at most
one pair of zero modes. Indeed, in a chiral basis, the nonzero off-diagonal  block of the
Dirac matrix is a square anti-symmetric matrix,
and generically has
one zero mode if the matrix is odd-dimensional and no zero modes if the
matrix is even dimensional.
In  Ref.~\cite{smilga}, in the sector of topological charge $k=0,\ldots,N_{\rm c}-1$, a total of  $2k(N_{\rm c}-k)$ zero modes are found, half of them right-handed and
the other half left-handed. However, these zero modes are only obtained
after complexifying the $\SU(N_{\rm c})$ algebra and are irrelevant in the
present context.
The corresponding  random matrix theory
for this universality class also has an anti-symmetric off-diagonal block
with no zero modes or one zero mode.

The joint probability density of the eigenvalues
is given by the form~\eqref{joint-chi} where the level repulsion is $|\lambda_a-\lambda_b|^4$ since all eigenvalues are Kramers degenerate
 (because the anti-unitary symmetry operator satisfies ($\sigma_2K)^2 = -1$). 
We rederive this joint probability density in appendix~\ref{app1.b} 
and relate it to the QCD Dirac operator in the microscopic limit. 
The repulsion of the eigenvalues from the origin is either linear 
($\alpha_{\rm D}=1$) for an even dimensional $W$ or quintic ($\alpha_{\rm D}=5$) for 
an odd-dimensional $W$. We emphasize that the pair of zero modes for odd-dimensional matrices is not related to topology.

\subsection{Symmetry Breaking Pattern} \label{sec:cont.3}
 
In table~\ref{table1}, we also summarize the symmetry breaking patterns for continuum QCD in two, three, and four dimensions (see~\cite{imbo} for a discussion of
general dimensions). We recall the results for the cases considered in our work and show that they also apply to the random matrix ensembles introduced in the previous subsection. We restrict ourselves
 to the two-dimensional case with the Dyson index $\beta_{\rm D}=1,4$. The other symmetry breaking patterns and their relation to random matrix theory were extensively discussed in Refs.~\cite{V,VZ}.

For $\beta_{\rm D} =1$, the off-diagonal block is symmetric after a unitary transformation,
$(\tau_2\mathcal{W}^{(2)})^T =\tau_2\mathcal{W}^{(2)}$ . Then we have
\be\label{Lag-int-bet1.a}
\bar \psi_{\rm R} \tau_2\mathcal{W}^{(2)} \psi_{\rm R} = \frac{1}{2} \bvec \bar \psi_{\rm R}^T \\ \psi_{\rm R} \evec^T
\bmat 0 & \tau_2\mathcal{W}^{(2)} \\ -\tau_2\mathcal{W}^{(2)} & 0 \emat \bvec \bar \psi_{\rm R}^T \\ \psi_{\rm R} \evec,
\ee
where $\psi_{\rm R}=(\eins+\sigma_3)\psi/2$ is the right handed component of a quark field $\psi$.  We obtain a similar expression for the other off-diagonal block, $\mathcal{W}^{(2)\,\dagger}$, of the  Dirac operator $\mathcal{D}^{(2)}$ with $\psi_{\rm R} \to \psi_{\rm L}=(\eins-\sigma_3)\psi/2$, i.e.
\be\label{Lag-int-bet1.b}
\bar \psi_{\rm L} (\tau_2\mathcal{W}^{(2)})^\dagger \psi_{\rm L} 
= \frac{1}{2} \bvec \bar \psi_{\rm L}^T \\ \psi_{\rm L} \evec^T
\bmat 0 & (\tau_2\mathcal{W}^{(2)})^\dagger \\ -(\tau_2\mathcal{W}^{(2)})^\dagger& 0 \emat 
\bvec \bar \psi_{\rm L}^T \\ \psi_{\rm L} \evec,
\ee
Therefore, the chiral symmetry is
$\USp(2N_{\rm f})\times \USp(2N_{\rm f})$ and acts on the doublets via the transformation $(\bar\psi_{\rm R}, \psi_{\rm R}^T)\to (\bar\psi_{\rm R}, \psi_{\rm R}^T)U_{\rm R}$ and $(\bar\psi_{\rm L}, \psi_{\rm L}^T)\to (\bar\psi_{\rm L}, \psi_{\rm L})^TU_{\rm L}$ with $U_{\rm R/L}\in\USp(2N_{\rm f})$. In terms of these doublets the chiral condensate can be written as
\be\label{cond-bet1}
\bar \psi_{\rm R} \psi_{\rm L}+\bar \psi_{\rm L} \psi_{\rm R}
=\bvec \bar \psi_{\rm R}^T \\ \psi_{\rm R} \evec^T
\bmat 0 & \eins \\ -\eins & 0 \emat \bvec \bar \psi_{\rm L}^T \\ \psi_{\rm L} \evec.
\ee
A non-zero expectation value  of the chiral condensate
requires that  the unitary symplectic matrices  fulfill the constraint
\be\label{symbreak-bet1}
 U_{\rm R}\bmat 0 & \eins \\ -\eins & 0 \emat U_{\rm L}^T=\bmat 0 & \eins \\ -\eins & 0 \emat ,
\ee
so that the chiral symmetry is broken to $\USp(2N_{\rm f}).$

This result can be derived by an explicit calculation for the corresponding
random matrix model, see appendix~\ref{app1.a.b}, and was also found in Ref.~\cite{imbo} for general QCD-like theories and in Ref.~\cite{class} for random matrix theories.

For two dimensional QCD with adjoint fermions ($\beta_{\rm D} =4$) we have that $\mathcal{W}^{(2)\,T} =-W^{(2)}$ is anti-symmetric so that the coupling of the gauge fields and  the quarks can be rewritten as
\be\label{Lag-int-bet4.a}
\bar \psi_{\rm R} \mathcal{W}^{(2)} \psi_{\rm R} = \frac{1}{2} \bvec \bar \psi_{\rm R}^T \\ \psi_{\rm R} \evec^T
\bmat 0 & \mathcal{W}^{(2)} \\ \mathcal{W}^{(2)} & 0 \emat \bvec \bar \psi_{\rm R}^T \\ \psi_{\rm R} \evec,
\ee
and
\be\label{Lag-int-bet4.b}
\bar \psi_{\rm L} \mathcal{W}^{(2)\,\dagger} \psi_{\rm L} = \frac{1}{2} \bvec \bar \psi_{\rm L}^T \\ \psi_{\rm L} \evec^T
\bmat 0 & \mathcal{W}^{(2)\,\dagger} \\ \mathcal{W}^{(2)\,\dagger} & 0 \emat \bvec \bar \psi_{\rm L}^T \\ \psi_{\rm L} \evec.
\ee
The corresponding chiral symmetry is ${\rm O}(2N_{\rm f})\times {\rm O}(2N_{\rm f})$ with the transformation $(\bar\psi_{\rm R}, \psi_{\rm R}^T)\to (\bar\psi_{\rm R}, \psi_{\rm R}^T)AO_{\rm R}A^{-1}$ and $(\bar\psi_{\rm L}, \psi_{\rm L}^T)\to (\bar\psi_{\rm L}, \psi_{\rm L}^T)AO_{\rm L}A^{-1}$ with $O_{\rm R/L}\in{\rm O}(2N_{\rm f})$ and
\be\label{A-def}
A^T \bmat 0 & \eins \\ \eins & 0 \emat A=\bmat \eins &  0 \\  0 & \eins \emat.
\ee
Invariance of the non-zero chiral condensate,
\be\label{cond-bet2}
\bar \psi_{\rm R} \psi_{\rm L}+\bar \psi_{\rm L} \psi_{\rm R}
=\bvec \bar \psi_{\rm R}^T \\ \psi_{\rm R} \evec^T
\bmat 0 & \eins \\ -\eins & 0 \emat \bvec \bar \psi_{\rm L}^T \\ \psi_{\rm L} \evec,
\ee
requires
\be
O_{\rm R}  = \bmat 0 &\eins \\ -\eins & 0\emat O_{\rm L}
\bmat 0 &\eins \\ -\eins & 0\emat ,
\ee
such that the symmetry is broken to ${\rm O}(2N_{\rm f})$. Also this case agrees with results of Refs.~\cite{class,imbo}.

\section{Two Dimensional Lattice QCD with Naive Fermions at Strong Coupling}\label{sec3}

In this section we consider the microscopic limit of naive fermions in 
the strong coupling limit and the corresponding random matrix theories. 
Thus the links, the gauge group elements on the lattice, are distributed according to  
the Haar-measure of the gauge group. In Secs.~\ref{sec3.1} we discuss the general effect of the parity of the lattice on the global symmetries of the Dirac operator. 
This discussion is combined with the specific anti-unitary symmetries of the QCD-like theories in Secs.~\ref{sec3.2}, \ref{sec3.3}, and \ref{sec3.4}. In  particular, 
we classify  each lattice Dirac operator according to a 
random matrix ensemble, which is summarize in table~\ref{table2} together with some spectral properties. These random matrix theory predictions are compared with 2-dim lattice simulations confirming that the parity of the
lattice has an important effect on the properties of the smallest eigenvalues. This was observed before in the  condensed matter literature \cite{markos}.

\subsection{General lattice model}\label{sec3.1}

The covariant derivatives that enter in the lattice QCD Dirac operator
can be readily constructed via the translation matrices. 
Before  doing so we introduce the lattice. Let $|j\rangle$ be the $j$'th site 
in one direction of a lattice written in Dirac's bra-ket notation. Then the 
dual vector is $\langle j|$. The translation matrices of an $L_1\times L_2$ lattice in the directions $\mu=1,2$ are given by
\begin{eqnarray}\label{3.1.1}
T_\mu=\left\{\begin{array}{cl} \displaystyle\underset{1\leq j\leq L_2}{\underset{1\leq i\leq L_1}{\sum}} 
U_{1 ij}\otimes|i\rangle\langle i+1|\otimes|j\rangle\langle j|, & \mu=1, 
\\ \displaystyle\underset{1\leq j\leq L_2}{\underset{1\leq i\leq L_1}{\sum}} 
U_{2 ij}\otimes|i\rangle\langle i|\otimes|j\rangle\langle j+1|, & \mu=2. \end{array}\right.
\end{eqnarray}
The matrices $U_{\mu ij}$ are given in some representation of 
the special unitary group ${\rm SU}(N_{\rm c})$ and are weighted with the Haar-measure of ${\rm SU}(N_{\rm c})$. 
Hence, the translation matrices $T_\mu$ are unitary. 

Note that our lattices have a toroidal geometry.  We have numerically looked at the effect of periodic and anti-periodic fermionic boundary conditions on the spectrum of the Dirac operator. Indeed, the universality class remains unaffected since the global symmetries are independent of the boundary conditions. Only the Thouless energy marginally changes.

The Dirac operator on a two dimensional lattice is given by
\begin{eqnarray}\label{3.1.2}
 D&=&\sigma_\mu(T_\mu -T_\mu^\dagger)\nn \\
&=&
\left[\begin{array}{cc} 0 & W \\ -W^\dagger & 0 \end{array}\right]=\left[\begin{array}{cc} 0 & T_x-T_x^\dagger+\imath(T_y-T_y^\dagger) \\ T_x-T_x^\dagger-\imath(T_y-T_y^\dagger) & 0 \end{array}\right].
\end{eqnarray}
Due to the lattice structure, an additional symmetry can exist in each direction 
if the number of  sites in a direction is even. Then the matrix elements of the 
Dirac operator between even and odd sites are non-vanishing while there is no direct coupling between an even and an even lattice site and between 
 an odd and an  odd site. For a two dimensional lattice we can  distinguish
three cases. First, the number of lattice sites $L_1$ and $L_2$ are both odd. Then, there are no additional symmetries such that the lattice Dirac operator is in the same symmetry class as the continuum theory. 
The other two cases are, second, $L_1$ even and $L_2$  odd or the reverse,  and third,  both  $L_1$ and $L_2$  are even.
We analyze these two cases in detail for each anti-unitary symmetry class separately.
Thereby we assume that both $L_1$ and $L_2$ are larger than $2$ because only then the low-lying eigenvalues
of the Dirac operator show a generic behavior. 

Let us define the operators
\begin{eqnarray}\label{3.1.3}
\Gamma_5^{(\mu)}=\left\{\begin{array}{cl} \displaystyle\underset{1\leq j\leq L_2}{\underset{1\leq i\leq L_1}{\sum}} (-1)^{i}\eins_{N_{\rm c}}\otimes|i\rangle\langle i|\otimes|j\rangle\langle j|, & \mu=1, \\ \displaystyle\underset{1\leq j\leq L_2}{\underset{1\leq i\leq L_1}{\sum}} (-1)^{j}\eins_{N_{\rm c}}\otimes|i\rangle\langle i|\otimes|j\rangle\langle j|, & \mu=2. \end{array}\right.
\end{eqnarray}
Then one can  show that the operator $\Gamma_5^{(\mu)}$ fulfills the  relation
\begin{eqnarray}\label{3.1.4}
\Gamma_5^{(\mu)} T_\omega \Gamma_5^{(\mu)}=(-1)^{\delta_{\mu\omega}}T_\omega
\end{eqnarray}
if $L_\mu$ is even. Hereby we employ the Kronecker symbol $\delta_{\mu\omega}$ in the exponent of the sign.

Let us consider the simplest case where $L_1$ and $L_2$ are odd. Then $W$ has no additional symmetries resulting from the lattice structure. Therefore, the Dirac operator will have the same unitary and anti-unitary symmetries
as in the continuum limit discussed in section~\ref{sec:cont}, in particular it is anti-Hermitian and chirally symmetric,
\begin{equation}\label{symgenlat1}
 D=-D^\dagger\quad{\rm and}\quad [\sigma_3,D]_+=0.
\end{equation}
Therefore the Dirac operator has the structure
\begin{equation}\label{formgenlat1}
 D=\left(\begin{array}{cc} 0 & W \\ -W^\dagger & 0 \end{array}\right),
\end{equation}
where $W$ may fulfill some additional anti-unitary symmetries because of the representation of the gauge theory.

In the second case, we have in one direction an even number of lattice sites and in the other direction an odd number of lattice sites. Let us assume that without loss of generality $L_1\in2\mathbb{N}$ and $L_2\in2\mathbb{N}+1$. Then the lattice Dirac operator fulfills the global symmetries
\begin{equation}\label{symgenlat2}
 D=-D^\dagger,\quad [\sigma_3,D]_+=0,\quad{\rm and}\quad[ \Gamma_5^{(1)}\sigma_2,D]_-=0
\end{equation}
plus possible anti-unitary symmetries depending on the representation of the gauge group. From the first two symmetries 
it follows that the Dirac operator has the chiral structure~\eqref{formgenlat1}. The last symmetry relation of Eq.~\eqref{symgenlat2} tells us hat the matrix $W$ is $\Gamma_5^{(1)}$-Hermitian, i.e.
\begin{equation}\label{relWgenlat2}
 W^\dagger=\Gamma_5^{(1)}W\Gamma_5^{(1)}.
\end{equation}
Hence the Dirac operator for this kind of lattices takes the form
\begin{equation}\label{formgenlat2}
 D=\left(\begin{array}{cc} 0 & \Gamma_5^{(1)}H \\ -H\Gamma_5^{(1)} & 0 \end{array}\right)=\diag(\Gamma_5^{(1)},\eins)\left(\begin{array}{cc} 0 & H \\ -H & 0 \end{array}\right)\diag(\Gamma_5^{(1)},\eins),
\end{equation}
with $H$ a Hermitian matrix. This matrix $H$ may be restricted to a subspace of the Hermitian matrices if we take into account the anti-unitary symmetries resulting from the representation of the gauge theory. The unitary matrix $\diag(\Gamma_5^{(1)},\eins)$ does not change the eigenvalue spectrum of $D$ and can be omitted.

One can also derive the structure~\eqref{formgenlat2} by employing the projection operators $(1\pm\Gamma_5^{(1)})/2$. They project the lattice onto sub-lattices associated to the even and odd lattice sites in the direction $\mu=1$. In such a basis, the translation matrix $T_1$ maps the even lattice sites to the odd ones and vice versa while the translation matrix $T_2$ maps the two sub-lattices onto themselves.

In the third case the lattice has 
an even number of lattice sites in both directions. This is exactly the situation of staggered fermions. The corresponding Dirac operator for naive fermions has the symmetries
\begin{equation}\label{symgenlat3}
 D=-D^\dagger,\quad [\sigma_3,D]_+=0,\quad[ \Gamma_5^{(1)}\sigma_2,D]_-=0,\quad{\rm and}\quad[ \Gamma_5^{(2)}\sigma_1,D]_-=0.
\end{equation}
Again the Dirac operator has the chiral structure~\eqref{formgenlat1}, but the symmetry relation of the matrix $W$ is given by
\begin{equation}\label{relWgenlat3}
 W^\dagger=\Gamma_5^{(1)}W\Gamma_5^{(1)}\quad{\rm and}\quad [\Gamma_5^{(1)}\Gamma_5^{(2)},W]_+=0.
\end{equation}
The first symmetry restricts $W$ to a $\Gamma_5^{(1)}$-Hermitian matrix 
whereas the second relation reflects the even-odd symmetry of
the Dirac operator. Therefore the lattice Dirac operator has the structure
\begin{equation}\label{formgenlat3}
 D=\diag(\Gamma_5^{(1)},\eins)\left(\begin{array}{c|c} 0 & \begin{array}{cc} 0 & X \\
 X^\dagger & 0 \end{array} \\ 
\hline \begin{array}{cc} 0 & -X \\ 
-X^\dagger & 0 \end{array} & 0 \end{array}\right)\diag(\Gamma_5^{(1)},\eins),
\end{equation}
where $X$ is a complex matrix  that may fulfill  anti-unitary symmetries depending on the representation of the gauge fields. The double degeneracy is immediate and is eliminated for staggered fermions.

Again one can also explicitly construct the form of the lattice Dirac operator~\eqref{formgenlat3} by employing the four projection operators $(1\pm\Gamma_5^{(1)})/2$ and $(1\pm\Gamma_5^{(2)})/2$. They split the lattice into four sub-lattices which are coupled via the translation matrices $T_{1/2}$.

Adding the anti-unitary symmetries  to the symmetries~\eqref{symgenlat1},~\eqref{symgenlat2}, and~\eqref{symgenlat3} will give rise to further constraints on $W$. In table~\ref{table2} we summarize these cases for each anti-unitary symmetry class. 
In general, the symmetry class will differ from the symmetry class in continuum. Therefore the corresponding random matrix ensemble and the symmetry breaking pattern
 will also change. In particular, one has to replace the indices $\beta_{\rm D}$ (Dyson index = level repulsion) and $\alpha_{\rm D}$ (=repulsion of the levels from the origin) in the joint probability densities of the eigenvalues of the random matrix model, cf. Eqs.~\eqref{joint-nonchi} and \eqref{joint-chi}, by effective values,
\begin{equation}\label{indeff}
 \beta_{\rm D}\to\beta_{\rm D}^{\rm (eff)}\quad{\rm and}\quad\alpha_{\rm D}\to\alpha_{\rm D}^{\rm (eff)}.
\end{equation}
This  impacts the spectral properties of the Dirac operator in the microscopic limit.

\begin{table}
\begin{tabular}{|c|c|c|c|c|c|c|c|c|}\hline
 Gauge theory & \ $\beta_{\rm D}$\ \ & Lat. & Sym. Class & $\beta_{\rm D}^{\rm (eff)}$&  $\alpha_{\rm eff}$ & Deg & ZM & Symmetry Breaking Pattern \\
\hline\hline
$N_{\rm c}=2$, fund. & 1   &   ee  & CII & 4&3 & 4& 0& $\U(4N_{\rm f}) \to \Ort(4N_{\rm f})$\\
$N_{\rm c}=2$, fund. & 1   &   eo  & C  &2 & 2 & 2 & 0 & $\USp(4N_{\rm f}) \to \U(2N_{\rm f})$     \\
$N_{\rm c}=2$, fund. & 1   & oo  & CI & 1&1 &1 & 0 & $\USp(2N_{\rm f}) \times \USp(2N_{\rm f}) \to \USp(2N_{\rm f})$\\
\hline
$N_{\rm c}>2$, fund. & 2   & ee  &AIII  & 2& 1 & 2& 0 & $\U(2N_{\rm f})\times \U(2N_{\rm f}) \to \U(2N_{\rm f})$\\
$N_{\rm c}>2$, fund. & 2   & eo  & A & 2& 0 & 1 & 0 & $\U(2N_{\rm f})\to \U(N_{\rm f}) \times \U(N_{\rm f})$ \\
$N_{\rm c}>2$, fund. & 2   & oo &AIII  & 2 & 1 & 1& 0 & $\U(N_{\rm f})\times \U(N_{\rm f}) \to \U(N_{\rm f})$ \\
\hline
$N_{\rm c}\geq2$, adj. & 4   & ee  & BDI & 1& 0 & 2& 0 & $\U(4N_{\rm f}) \to \USp(4N_{\rm f}) $ \\
$N_{\rm c}\geq2$, adj. & 4   & eo  & D  & 2&  0& 2 & 0 & $\Ort(4N_{\rm f}) \to \U(2N_{\rm f})$\\
$N_{\rm c}\in2\mathbb{N}+1$, adj. & 4   & oo  & DIII (even-dim)  & 4 & 1 &2 & 0 & $\Ort(2N_{\rm f}) \times \Ort(2N_{\rm f}) \to \Ort(2N_{\rm f})$\\
$N_{\rm c}\in2\mathbb{N}$, adj. & 4   & oo  & DIII (odd-dim)   & 4 & 5 & 2& 2 & $\Ort(2N_{\rm f}) \times \Ort(2N_{\rm f}) \to \Ort(2N_{\rm f})$ \\ \hline
\end{tabular}
\caption[]{Random matrix theories for the two-dimensional naive lattice QCD Dirac operator 
with gauge group 
listed in the first column.  The Dyson index $\beta_{\rm D}$ 
refers to the anti-unitary symmetry of the Dirac operator 
in the continuum. Because of additional symmetries 
the power of the
Vandermonde determinant, $\beta_{\rm D}^{\rm  (eff)}$, is generally different 
from the continuum theory and thus, the level repulsion as well. 
Moreover the repulsion of the levels from the origin, 
namely $\lambda^{\alpha_{\rm eff}}$, the generic degeneracy of the eigenvalues 
(third to last column, ``Deg''), and the number of generic zero modes (second to last column, ``ZM'')  generally change as well. The third column refers to whether $L_1$ or $L_2$ are
even (e) or odd (o). All discretizations are classified according to the ten-fold classification of random matrix
theories (fourth column) which share the same pattern of chiral symmetry breaking with the lattice QCD Dirac operator (we do not consider axial symmetry breaking). Notice that the symmetry breaking pattern and, therefore, the global symmetries  of the lattices where $L_1$ and $L_2$ are both odd is the same with the two-dimensional QCD Dirac operator in continuum, cf. table~\ref{table1}. 
\label{table2}}
\end{table}

There are additional conditions on the off-diagonal block $W$ of the lattice Dirac operator $D$ which are independent
 of the gauge configurations.
 For example the traces of $W$ satisfy the relations
\begin{eqnarray}
\label{Wreltrace}
 \tr W^2=\tr W^{2l+1}=0\quad {\rm and}\quad 
\tr WW^\dagger=
\left \{
\begin{array}{cl} 2N_{\rm c}L_1L_2, & {\rm fundamental\ fermions,} \\ 
2(N_{\rm c}^2-1)L_1L_2, & {\rm adjoint\ fermions} \end{array}\right .
\end{eqnarray}
with $l=0,1,2,\ldots$ such that $l\leq\min\{L_1,L_2\}/2-1$. They result from the fact that the translation matrices~\eqref{3.1.1} are unitary and have no diagonal elements. 
The conditions of the kind~\eqref{Wreltrace} are expected to have no influence on the microscopic spectrum in the limit of large matrices. Nevertheless, they may give rise to finite volume corrections which turn out to be particularly large
for  the simulations of $\SU(3)$ gauge theory with fermions 
in the fundamental representation and choosing $L_1$ even and $L_2$ odd, see subsection~\ref{sec3.4.eo}.
 The effect of such conditions can also be studied with random matrix theory and we do this for the simplest condition, namely that $W$ is traceless, i.e. $\tr W=0$.

\subsection{${\SU}(2)$ and fermions in the fundamental representation}\label{sec3.2}

When studying the two-color theory in its fundamental representation the translation matrix fulfills exactly the same anti-unitary symmetry as the covariant derivative in the continuum theory,
\begin{eqnarray}\label{symtrabet1}
[\imath T_\mu,\tau_2 K]_- =0,
\end{eqnarray}
cf. Eq. \eqref{1.8}. This symmetry carries over to the symmetry
\begin{eqnarray}\label{symDirbet1}
[\imath D,\tau_2\sigma_2 K]_- =0,
\end{eqnarray}
for the lattice Dirac operator meaning that there is always a gauge field independent basis where the Dirac operator appears real. However, as 
is the case in the continnuum theory, the symmetry~\eqref{symDirbet1} may
not commute
with the symmetries~\eqref{symgenlat1}, \eqref{symgenlat2}, and \eqref{symgenlat3}. In the continuum theory we showed that the anti-unitary
symmetry  resuts in a symmetry of the off-diagonal block $W$,
\be\label{symmetry-lat-bet1b}
W= -\tau_2 W^T \tau_2.
\ee
cf. Eq.~\eqref{symmetry-cont-bet1b}. This carries over to the lattice  theory as well and  together  with the symmetries~\eqref{symgenlat1}, \eqref{symgenlat2}, and \eqref{symgenlat3} yields the symmetry classification given in Table 2.
This is worked out in detail in the subsections~\ref{sec3.2.oo}, \ref{sec3.2.oe}, and \ref{sec3.2.ee} for $(L_1,L_2)$ odd-odd, even-odd, and even-even, respectively.

\subsubsection{The Odd-Odd Case}\label{sec3.2.oo}

As already discussed before, this case does not have any additional symmetries and the pattern of chiral
symmetry breaking as well as the distribution of the eigenvalues in the microscopic domain has to be the same
as in the continuum limit which was discussed in section~\ref{sec3.1}. The symmetries of the Dirac operator are summarized in Eq.~\eqref{symm-bet1-2d} which 
translates in terms of the lattice Dirac operator as 
in Eqs.~\eqref{symgenlat1} and \eqref{symmetry-lat-bet1b}. 
That corresponds to a chiral random matrix theory with symmetric complex
off-diagonal blocks. In the Cartan classification of symmetric spaces,
this is denoted by the symbol CI.
 The corresponding microscopic level density is given by ($x=\lambda V\Sigma$) \cite{ivanov}
\begin{eqnarray}\label{3.2.mic.1}
\rho(x)=\frac{x}{2}\left[J_1^2(x)-J_{0}(x)J_2(x)\right]+\frac{1}{2}J_0(x)J_1(x).
\end{eqnarray}
The symmetry breaking pattern is therefore the same as in the continuum, namely $\U(2N_{\rm f})\to\Ort(2N_{\rm f})$.

\begin{figure}[!ht]
\centerline{\includegraphics[width=0.5\textwidth,clip=]{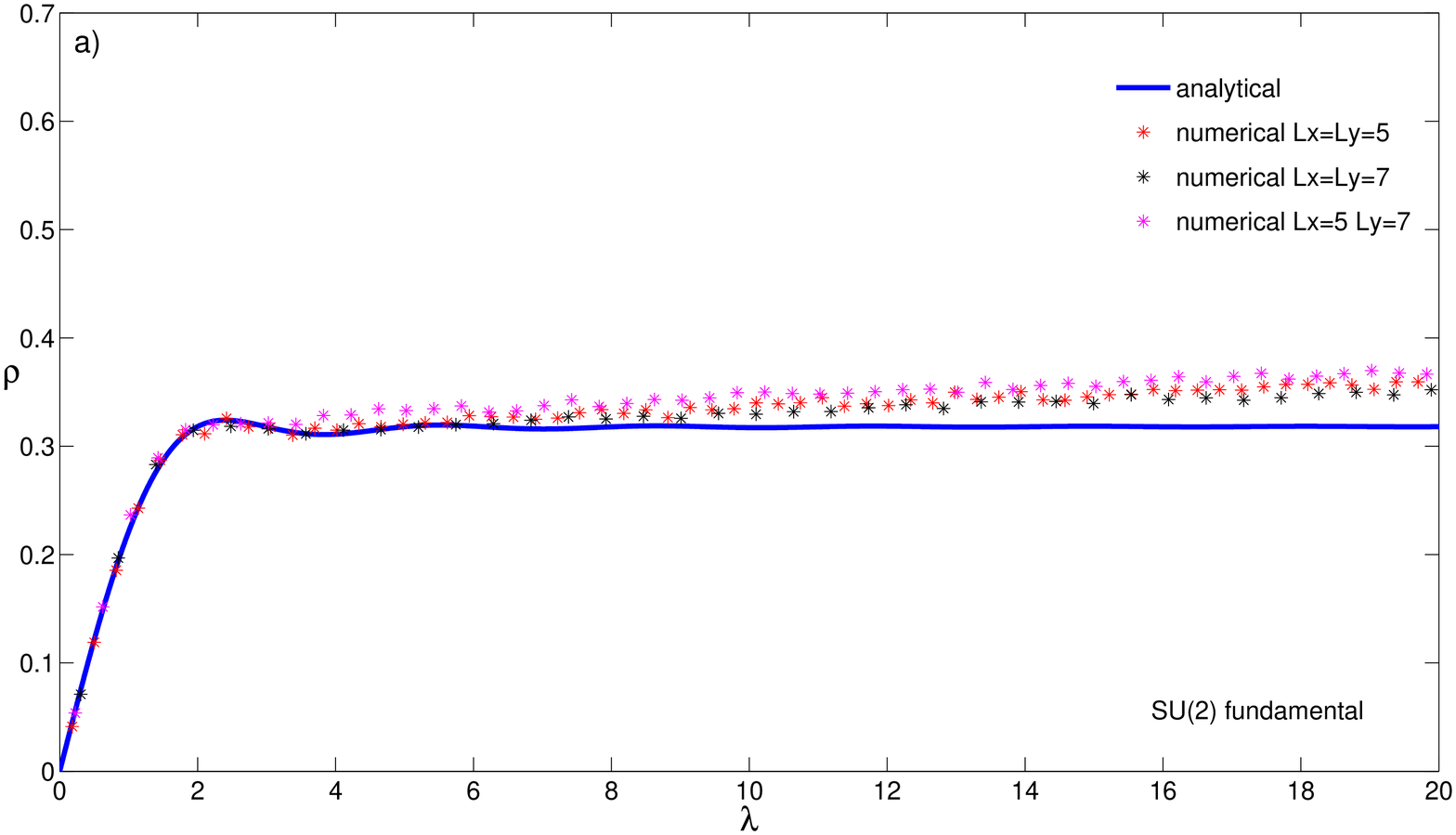}
\includegraphics[width=0.5\textwidth,clip=]{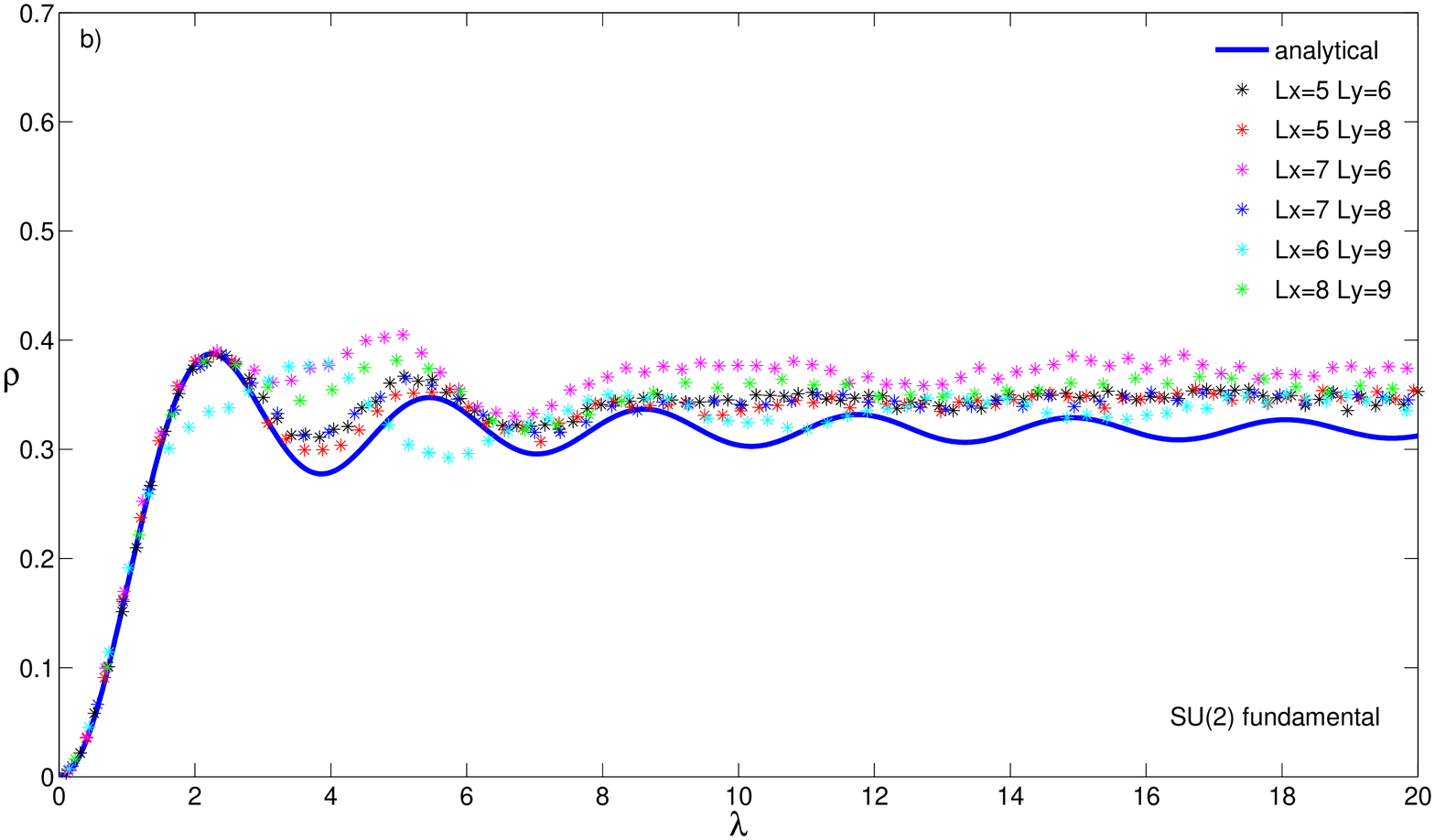}}
\centerline{\includegraphics[width=0.5\textwidth,clip=]{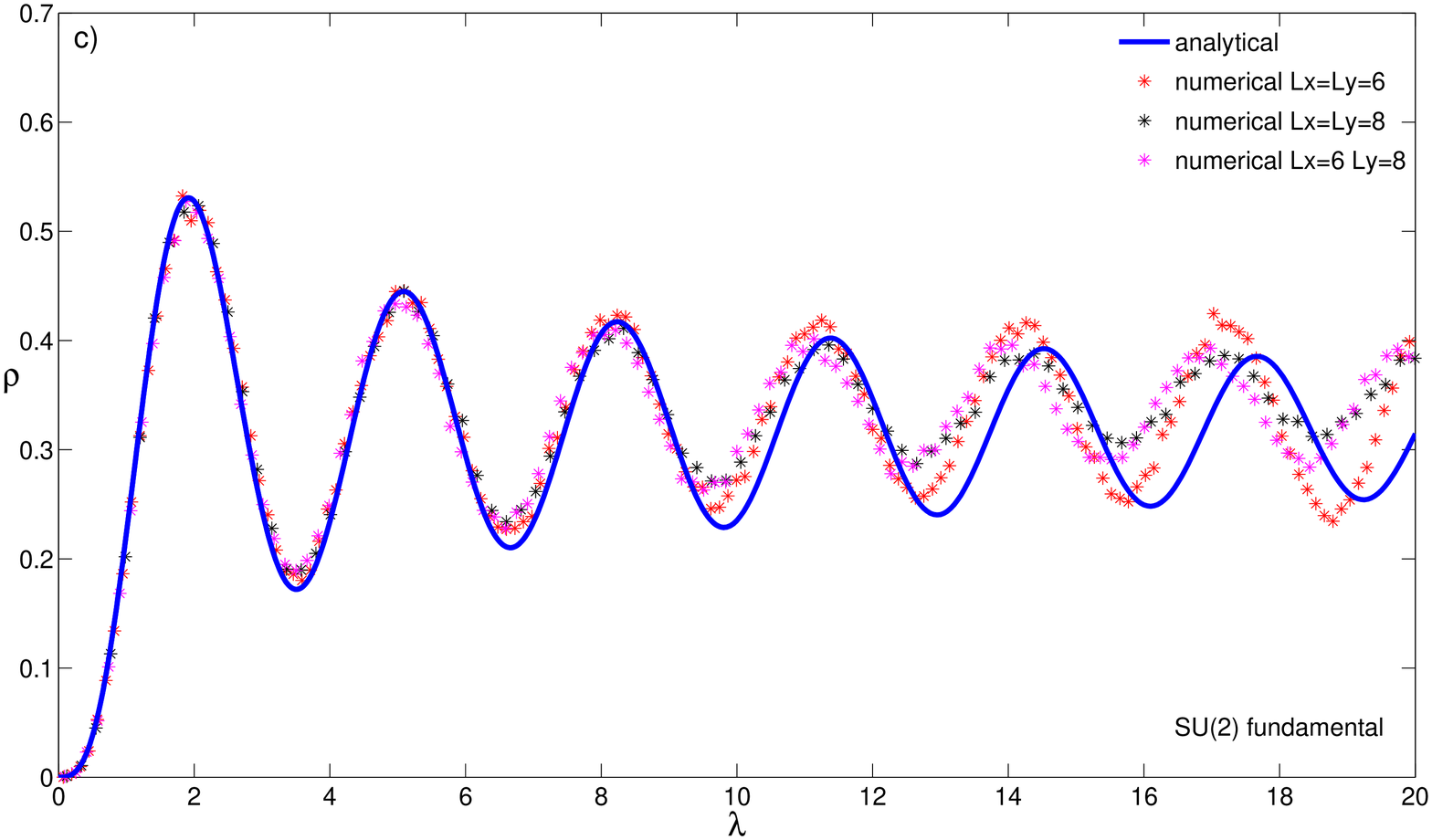}}
\caption{ Comparison of the microscopic level densities  of lattice QCD data in the strong coupling limit at various lattice sizes (stars) and the analytical results given by the corresponding random matrix theories (solid curves). The plotted lattice gauge theories are: a) $\SU(2)$ fundamental and $L_1,L_2$ odd, b) $\SU(2)$ fundamental
 and $L_1+L_2$ odd, and c) $\SU(2)$ fundamental
and $L_1,L_2$ even.}
\label{fig1}
\end{figure}

In Fig.~\ref{fig1}a  we compare the prediction~\eqref{3.2.mic.1}
for the 
low-lying Dirac spectrum  with lattice QCD data at strong coupling. The size of
the lattices is quite small. Nevertheless the agreement of the analytical 
prediction for the microscopic level density and the simulations around the origin is good.
 In particular, the linear repulsion of the eigenvalues from the origin is confirmed.
Also the degree of degeneracy and the number of generic zero modes, which are in this case one and zero, respectively, are confirmed.
The lattice results are obtained from an ensemble of  about $10^5$
independent configurations with the links generated by the Haar measure of the gauge group $\SU(2)$.

\subsubsection{The Even-Odd Case}\label{sec3.2.oe}

For definiteness we choose $L_1$  even and $L_2$  odd. 
Then, the Dirac operator is of  the form~\eqref{formgenlat2}. We combine the intermediate result $W=\Gamma_5^{(1)}H$ with a Hermitian matrix $H$ and the anti-unitary symmetry 
\eqref{symmetry-lat-bet1b}. Therefore we can find a gauge field independent rotation, namely $U_5^{(1)}=\exp[\pi\imath(\Gamma_5^{(1)}-\eins_{N_{\rm c}L_1L_2})/4]$, where $\widetilde{H}=U_5^{(1)}HU_5^{(1)\,-1}$ becomes an anti-self-dual  Hermitian matrix ($\widetilde{H}=\widetilde{H}^\dagger=-\tau_2\widetilde{H}^T\tau_2=-\tau_2\widetilde{H}^*\tau_2$). This is the class $C$ of the tenfold classification~\cite{class} and $\widetilde{H}$ is an element in the Lie-algebra of the group $\USp(N_{\rm c}L_1L_2)$.
In this basis,  the Dirac operator reads 
\begin{equation}\label{Dirac-lat-bet1-eo}
 D=\diag(U_5^{(1)},U_5^{(1)\,-1})\left(\begin{array}{cc} 0 & \widetilde{H} \\ -\widetilde{H} & 0 \end{array}\right)\diag(U_5^{(1)\,-1},U_5^{(1)}).
\end{equation}
Note that $\Gamma_5^{(1)}=U_5^{(1)\,2}$.

What does this imply for the spectrum of the Dirac operator? The anti-unitary symmetry leads to a pair of eigenvalues $\pm \lambda$ of the Hermitian matrix $\widetilde{H}$. Indeed, if $\lambda$ is an eigenvalue of $\widetilde{H}$ with the eigenvector $|\phi\rangle$,
\be\label{h3.2.a}
\widetilde{H}|\phi\rangle = \lambda|\phi\rangle,
\ee
then the state $\tau_2|\phi^*\rangle$ is an eigenvector with  eigenvalue $-\lambda$,
\be \label{h3.2.b}
\widetilde{H}\tau_2|\phi^*\rangle = -\lambda\tau_2|\phi^*\rangle.
\ee
Therefore the Dirac operator~\eqref{Dirac-lat-bet1-eo} has the eigenvalues $\pm\imath\lambda$ which are doubly degenerate. This leads to a doubling of the number of flavors and the spectrum of $D$ is twice the spectrum of $\imath\widetilde{H}$. In addition, because of 
\begin{eqnarray}\label{Lag-lat-bet1-eo}
\bar \psi^T \widetilde H \psi = \frac 12 (\bar \psi^T \widetilde H \psi
- (\tau_2 \psi)^T \widetilde H \tau_2\bar\psi),
\end{eqnarray}
the 
flavor symmetry is enhanced to $\USp(4N_{\rm f})$,
cf. Eq.~\eqref{Lag-int-bet1.a}. Because
\begin{eqnarray}\label{source-mass}
\det(D+m\eins) = \det(\widetilde H^2+m^2\eins)=\det(\widetilde H+\imath m\eins)\det(\widetilde H-\imath m\eins),
\end{eqnarray}
a nonzero eigenvalues density of $\widetilde H$ leads to a nonzero eigenvalue
density of the Dirac operator, $D$. 
The symmetry $\USp(4N_{\rm f})$ is thus spontaneously broken by 
the formation of a condensate with $m$ as source term.
 However this condensate is still invariant under a $\U(2N_{\rm f})$  subgroup of $\USp(4N_{\rm f})$
\be\label{symrelbet1}
[\diag(U,U^*)]^T\bmat 0 & m\eins_{2N_{\rm f}} \\ -m \eins_{2N_{\rm f}} & 0 \emat  
\diag(U,U^*) =\bmat 0 & m\eins_{2N_{\rm f}} \\ -m \eins_{2N_{\rm f}} & 0 \emat.  
\ee
 Thus the symmetry breaking pattern is $\USp(4N_{\rm f})\to\U(2N_{\rm f})$ in 
agreement with the symmetry breaking pattern of the corresponding random
matrix ensemble \cite{class}.

The joint probability distribution of the symmetry class C coincides with  the distribution of the non-zero eigenvalues of
 chGUE for $\nu = 1/2$.  
The microscopic level density is thus given by \cite{akemann-damgaard-jv,az,ivanov}
\begin{eqnarray}\label{3.2.mic.2}
\rho(x)=\frac{1}{\pi}-\frac{\sin(2x)}{2\pi x}.
\end{eqnarray}
In Fig. \ref{fig1}b we compare this result to lattice simulations. We find only good agreement to about one eigenvalue spacing. 
The reason for the strong disagreement above the average position of the first eigenvalue is not clear.
 Nevertheless, the quadratic repulsion of the eigenvalues from the origin, 
 the double degeneracy of the eigenvalues, and the fact that there are no generic zero modes are confirmed by the lattice simulations.

\subsubsection{The Even-Even Case}\label{sec3.2.ee}

Finally, we consider the case with both $L_1$ and $L_2$ even. 
Then the Dirac operator has the structure given in Eq.~\eqref{formgenlat3}.
 After combining the chiral structure of $W$ with the anti-unitary symmetry~\eqref{symmetry-lat-bet1b} the Dirac operator takes the form
\begin{equation}\label{Dirac-lat-bet1-ee}
 D=\diag(U_5^{(1)},U_5^{(1)\,-1})\left(\begin{array}{c|c} 0 & \begin{array}{cc} 0 & \widetilde{W} \\ \widetilde{W}^\dagger & 0 \end{array} \\ \hline \begin{array}{cc} 0 & -\widetilde{W} \\ -\widetilde{W}^\dagger & 0 \end{array} & 0 \end{array}\right)\diag(U_5^{(1)\,-1},U_5^{(1)})
\end{equation}
with $\widetilde{W}^*=\tau_2\widetilde{W} \tau_2$ a quaternion matrix without any further symmetries. The unitary transformation $\diag(U_5^{(1)\,-1},U_5^{(1)})$ is exactly the same as in the previous subsection and keeps the spectrum invariant such that the global symmetries of the lattice Dirac operator $D$ essentially coincide with the continuum Dirac operator in four dimensions with the fermions in the adjoint representation. Therefore, the random matrix ensemble
corresponding to this type  of lattice theory is the chGSE 
with the chiral symmetry breaking pattern $\U(4N_{\rm f})\to  \Ort(4N_{\rm f}) $. The degeneracy of the eigenvalues is four because of Kramers degeneracy and the doubling of flavors.
 In table~\ref{table2} we summarize the  main properties of this ensemble.

The microscopic level density of the lattice QCD Dirac operator in this class
is given by  the $\nu =0$ result of chGSE \cite{nagao,ivanov} 
(note that $\widetilde{W}$ is a square matrix),
\begin{eqnarray}\label{3.2.mic.3}
\rho(x)= x\left[J_0^2(2x)+J_1^2(2x)\right]-\frac{1}{2}J_{0}(2x)\int\limits_0^{2x}J_{0}(\widetilde{x})d\widetilde{x}.
\end{eqnarray} 
 There are no generic zero modes and the levels show a cubic repulsion from the origin.

In Fig.~\ref{fig1}c we compare the result~\eqref{3.2.mic.3} to lattice simulations  of the two-dimensional Dirac operator for QCD with two colors. There is an excellent agreement
for the first few eigenvalues confirming our predictions.

\subsection{${\rm SU}(N_{\rm c})$ and fermions in the adjoint representation.}\label{sec3.3}

For the fermions in the adjoint representation of the gauge group $\SU(N_{\rm c}\geq2)$ the translation matrices are real and, hence, satisfy the anti-unitary symmetry
\begin{eqnarray}\label{symtrabet4}
[K,T_\mu]_-=T_\mu.
\end{eqnarray}
On a $L_1 \times L_2$ lattice, the translation matrices are represented by a subset of matrices in the orthogonal group\\
 ${\rm O}((N_{\rm c}^2-1)L_1L_2)$. 
The symmetry~\eqref{symtrabet4} carries over to the two-dimensional 
lattice Dirac operator
\begin{eqnarray}\label{symDirbet4}
[\imath D,\sigma_2 K]_- =0,
\end{eqnarray} 
and its off-diagonal block matrix
\be\label{symmetry-lat-bet4b}
W= - W^T.
\ee
We combine this symmetry with the symmetries~\eqref{symgenlat1}, \eqref{symgenlat2}, and \eqref{symgenlat3} along the same lines as shown in subsection~\ref{sec3.3}. Thereby we discuss the odd-odd, even-odd, and even-even lattices in subsections~\ref{sec3.3.oo}, \ref{sec3.3.eo}, and \ref{sec3.3.ee} , respectively.

\subsubsection{The Odd-Odd Case}\label{sec3.3.oo}

In the case where both the number of lattice sites $L_1$ and $L_2$ are odd, the Dirac operator has the same
symmetries as in the continuum limit resulting in the same pattern of chiral
symmetry breaking ($\Ort(2N_{\rm f})\times\Ort(2N_{\rm f})\to\Ort(2N_{\rm f})$) and the same microscopic spectral properties (see table~\ref{table2}). Depending 
on the number of colors the off-diagonal  block $W$ of the
lattice Dirac operator is either even or odd dimensional and the corresponding symmetry class
is given by the second Bogolyubov-de Gennes ensemble DIII, see Ref.~\cite{class,az}, which can be also either even or odd, respectively.
The microscopic level density was obtained in  Ref.~\cite{ivanov} and is
given by
\begin{eqnarray}\label{3.3.mic.1a}
\rho(x)=\frac{x}{2}\left[2J_1^2(2x)+J_0^2(2x)-J_{0}(2x)J_2(2x)\right]+\frac{1}{2}J_1(2x)
\end{eqnarray}
for $N_{\rm c}$ odd and
\begin{eqnarray}\label{3.3.mic.1b}
\rho(x)=2\delta(x)+\frac{x}{2}\left[2J_1^2(2x)+J_0^2(2x)-J_{0}(2x)J_2(2x)\right]-\frac{1}{2}J_1(2x)
\end{eqnarray}
for $N_{\rm c}$ even.
Notice that the lattice Dirac operator has one additional pair 
of generic zero-modes if the number of colors is even otherwise 
there are no generic zero modes. Therefore the repulsion of 
the eigenvalues from the origin is stronger. However, the level repulsion is always quartic, see table~\ref{table2}. Moreover, the full spectrum is Kramers degenerate. This is a  characteristic for ensembles associated to the Dyson index $\beta_{\rm D}=4$.

\begin{figure}[!ht]
\centerline{\includegraphics[width=0.5\textwidth,clip=]{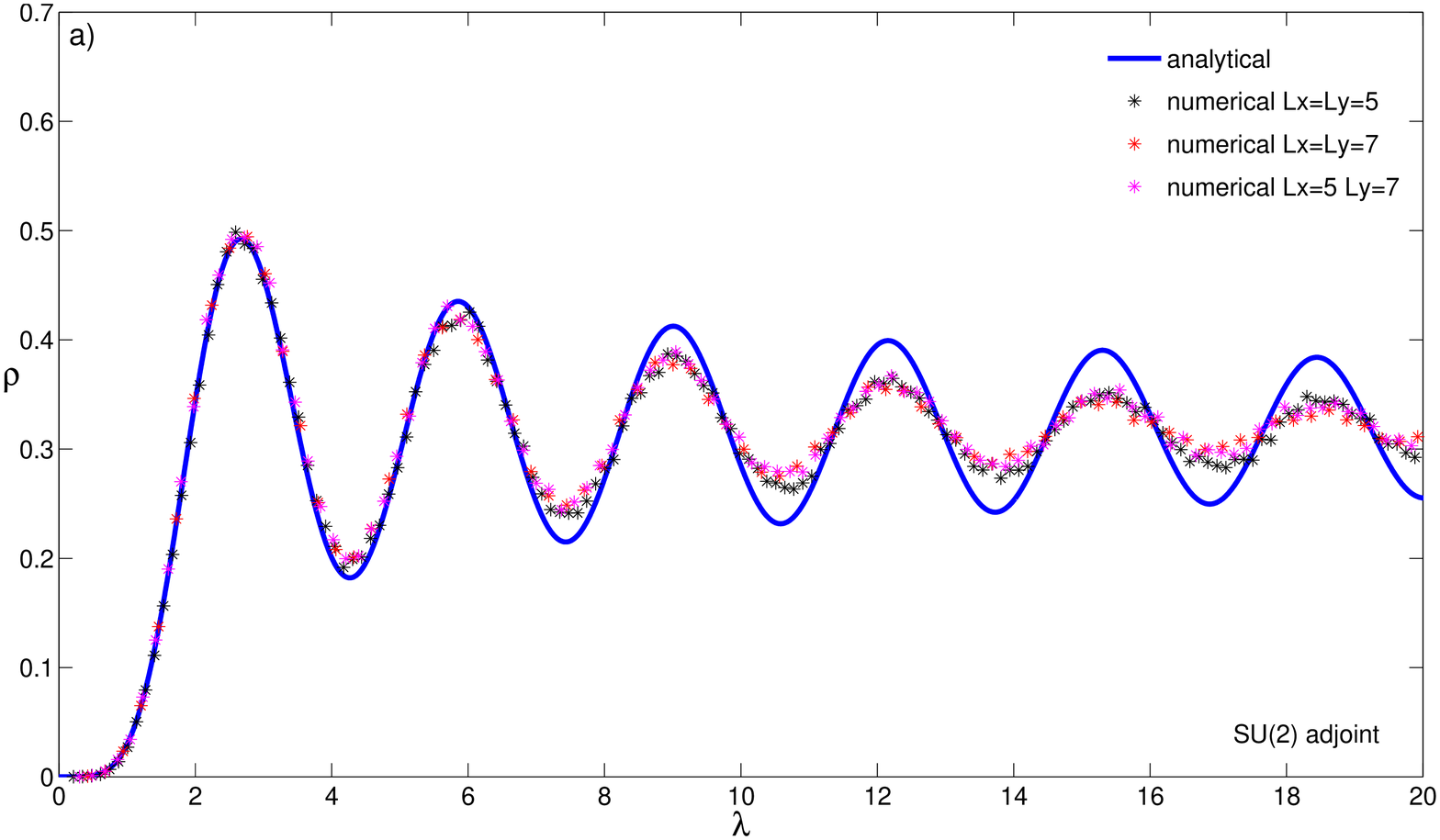}
\includegraphics[width=0.5\textwidth,clip=]{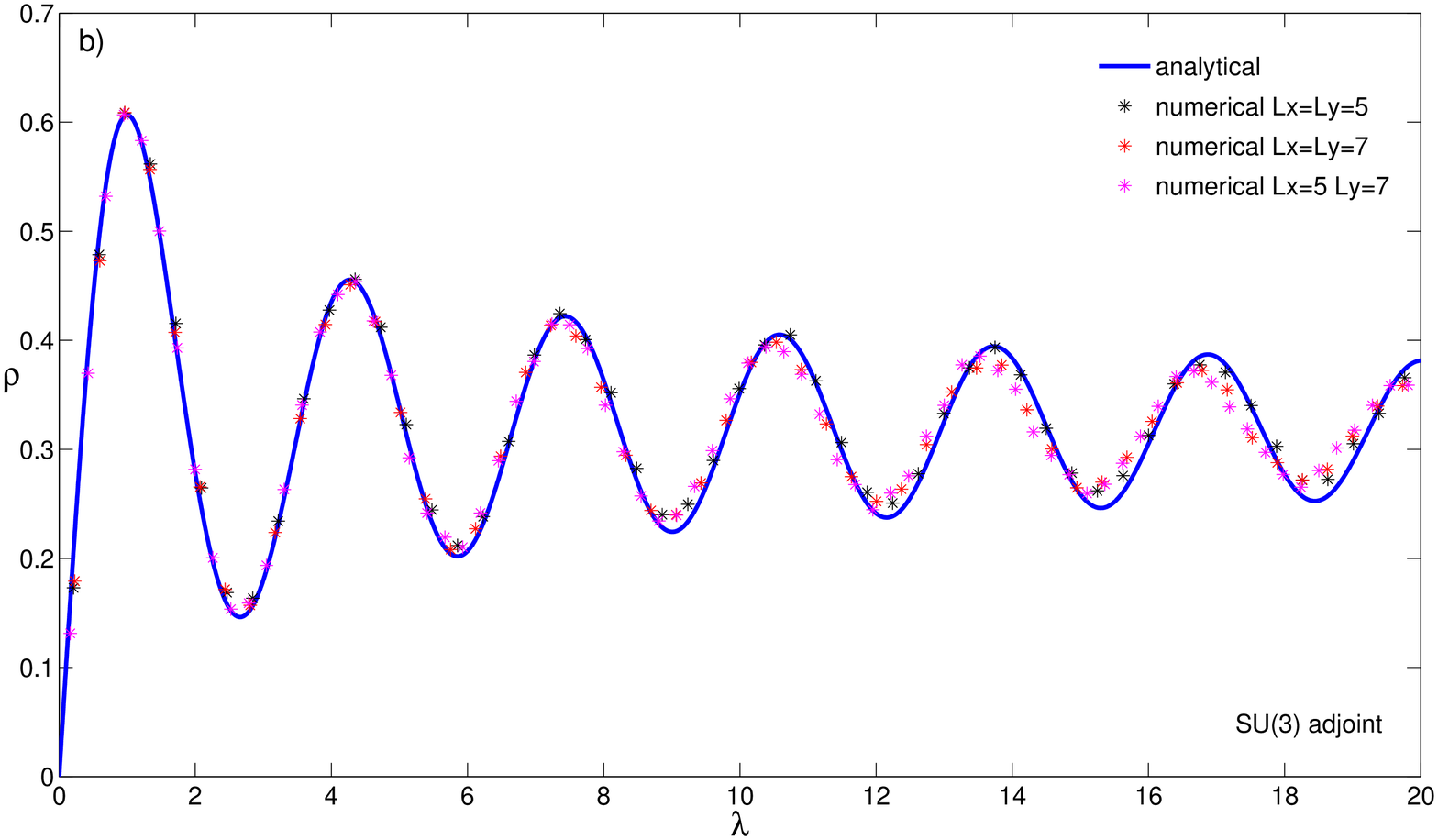}}
\centerline{\includegraphics[width=0.5\textwidth,clip=]{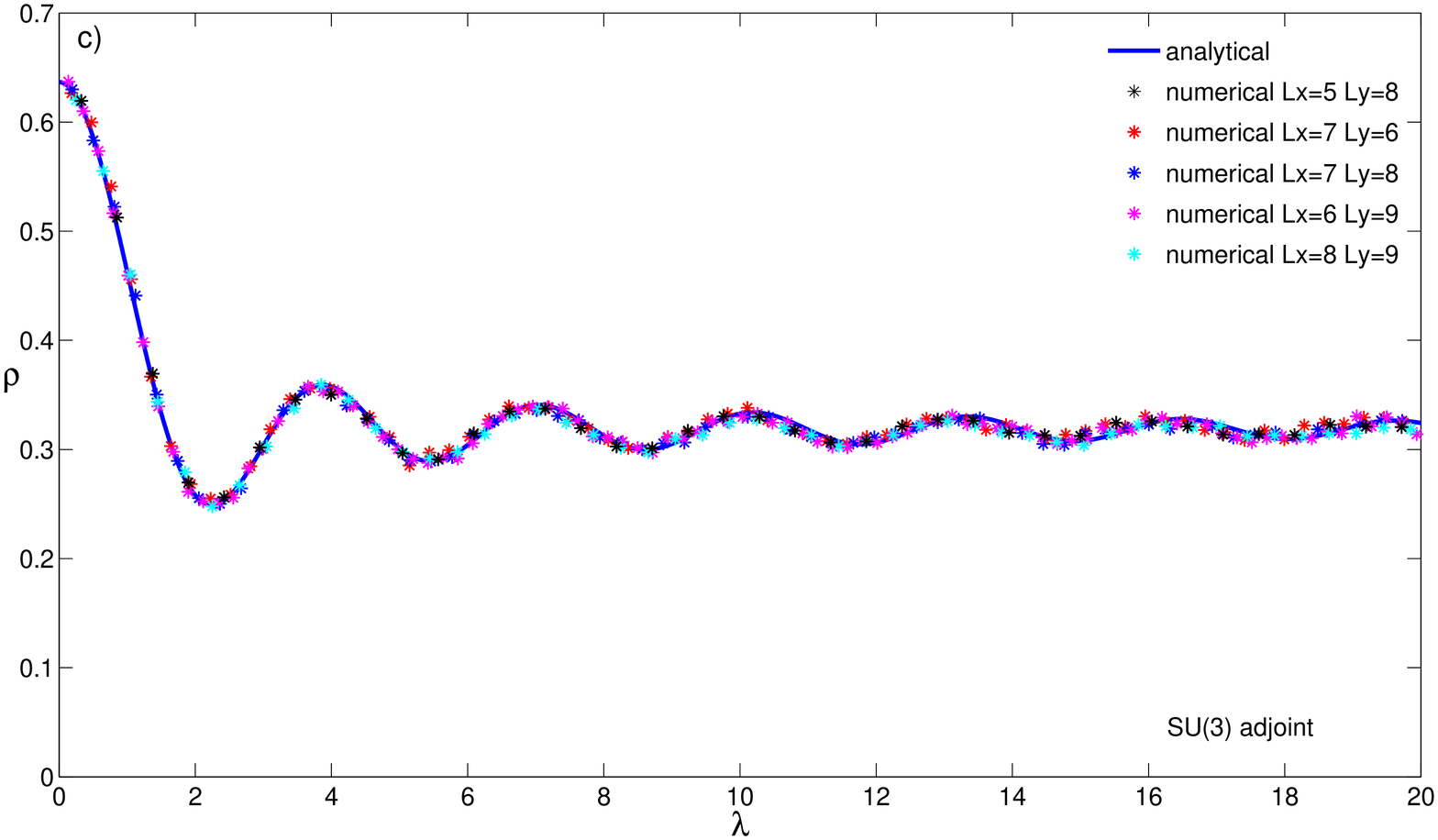}
\includegraphics[width=0.5\textwidth,clip=]{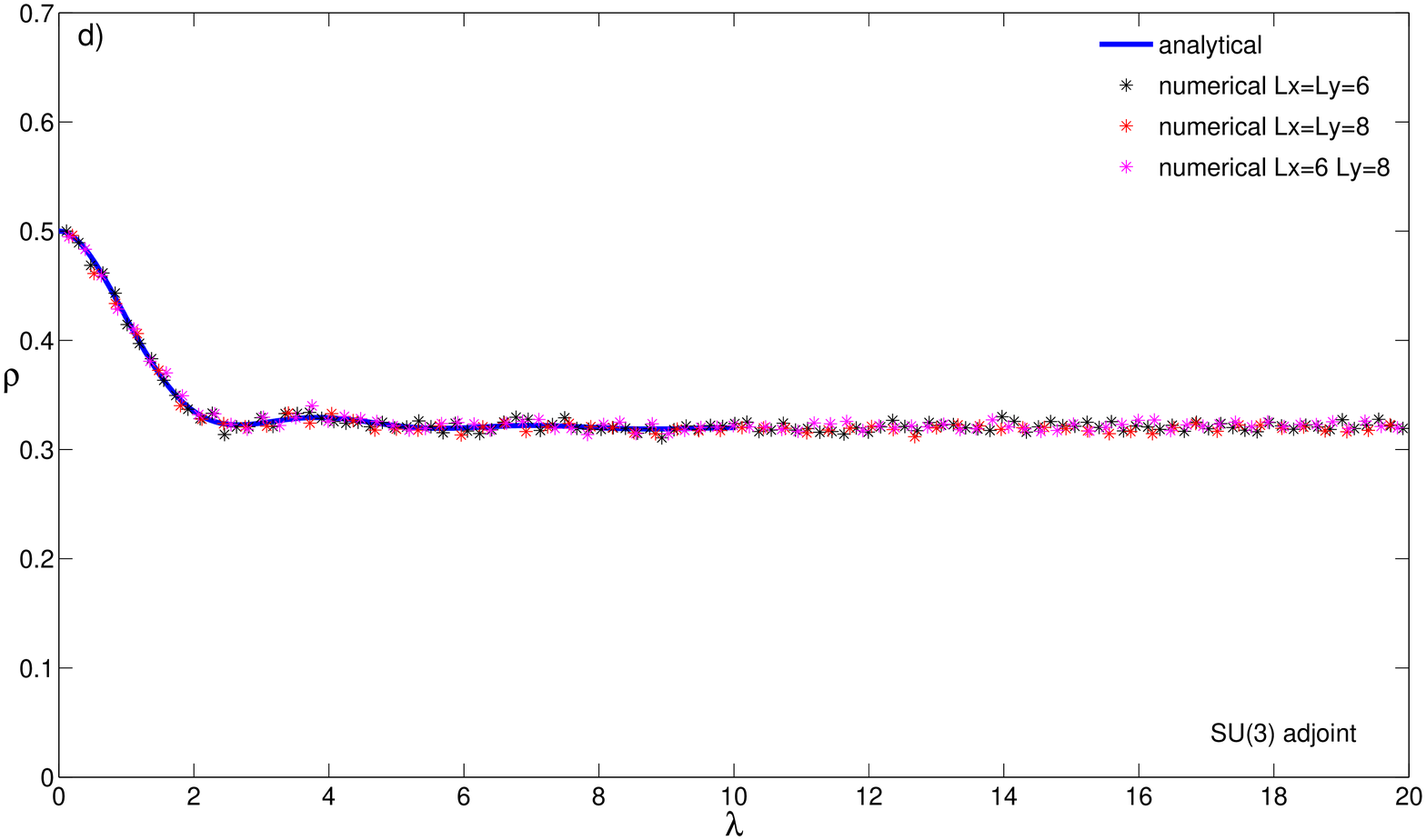}}
\caption{ Comparison of the microscopic level densities  
of the lattice QCD Dirac operator in the strong coupling limit at various lattice sizes (stars) and the analytical results derived from the corresponding random matrix theories (solid curves). Results are shown for the lattice
theories: a) $\SU(2)$ adjoint and $L_1,L_2$ odd, b) $\SU(3)$ adjoint and $L_1,L_2$ odd, c) $\SU(3)$ adjoint
 and $L_1$ odd and $L_2$ even, and d) $\SU(3)$ adjoint
and $L_1,L_2$ even.}
\label{fig2}
\end{figure}

In Figs.~\ref{fig2}a and \ref{fig2}b we compare the low lying lattice Dirac spectra and the analytical results of
~\eqref{3.3.mic.1a} and \eqref{3.3.mic.1b} for two and three colors, respectively. The agreement is good for the first few eigenvalues and becomes better when increasing the number of colors. 

\subsubsection{The Even-Odd Case}\label{sec3.3.eo}

Next we consider the mixed situation where the lattice has an even $L_1$ and an odd $L_2$. The combination of the symmetries~\eqref{symgenlat2} and \eqref{symmetry-lat-bet4b} can be again simplified via the same unitary transformation $\diag(U_5^{(1)\,-1},U_5^{(1)})$ as introduced in subsection~\ref{sec3.2.oe}. Then the lattice Dirac operator can be written as
\begin{equation}\label{Dirac-lat-bet4-eo}
 D=\diag(U_5^{(1)},U_5^{(1)\,-1})\left(\begin{array}{cc} 0 & \widetilde{H} \\ -\widetilde{H} & 0 \end{array}\right)\diag(U_5^{(1)\,-1},U_5^{(1)}),
\end{equation}
where $\widetilde{H}$ is purely imaginary and anti-symmetric. Thus the symmetry class is equivalent to a random matrix ensemble with the matrices in the Lie-algebra of the orthogonal group $\Ort(L_1L_2(N_{\rm c}^2-1))$ which is denoted by the Cartan symbol D \cite{class}. 
Although for this ensemble one also has to distinguish between even and odd matrix size $N$ because of  an additional pair of generic zero modes, the lattice Dirac operator  always yields an even sized matrix $\widetilde{H}$. The reason is that $\widetilde{H}$ is $L_1L_2(N_{\rm c}^2-1)\times L_1L_2(N_{\rm c}^2-1)$ dimensional where $L_1$ is even. 
Therefore we expect a quadratic level repulsion, no repulsion of the levels of $D$ from the origin and no generic zero modes, cf. table~\ref{table2}.
  The number of flavors is doubled because of the particular block structure~\eqref{Dirac-lat-bet4-eo}.

 The quark bilinear can be written as
\be 
\bar \psi^T \widetilde{H} \psi= 
\frac{1}{2} ( \bar \psi \widetilde{H} \psi 
+  \psi \widetilde{H} \bar \psi),
\ee
so that  the symmetry group is $\Ort(4N_{\rm f})$.  As  was shown in the 
case $\beta_{\rm D}=1$, see subsection~\ref{sec3.2.oe}, a nonzero eigenvalue
density of $\widetilde{H}$ results in a nonzero eigenvalues density of the
Dirac operator resulting in a chiral condensate with source term $m$.
This condensate breaks the $\Ort(4N_{\rm f})$ symmetry group to
the subgroup satisfying
\be\label{symrelbet4}
O^T\bmat 0 & m\eins_{2N_{\rm f}} \\ -m \eins_{2N_{\rm f}} & 0 \emat  
O =\bmat 0 & m\eins_{2N_{\rm f}} \\ -m \eins_{2N_{\rm f}} & 0 \emat 
\ee
 This equation enforces the matrix $O$ to a block structure
\be\label{symrelbet4.b}
O =\bmat O_1 & O_2 \\ -O_2 & O_1 \emat.  
\ee
The orthogonality of $O$ requires that
\be\label{symrelbet4.c}
(O_1+iO_2)^\dagger (O_1+iO_2) = \eins
\ee
so that $O$ is equivalent to a unitary transformation. Moreover each unitary matrix $U\in\U(2N_{\rm f})$ can be decomposed into the real matrices $O_1 = \frac 12(U+U^*)$ and $O_2 =-i(U-U^*)$. 
 Hence the remaining group invariance is equal to $\U(2N_{\rm f})$ yielding the symmetry breaking pattern $\Ort(4N_{\rm f})\to\U(2N_{\rm f})$.

The microscopic level density can be calculated from 
the corresponding random matrix ensemble in class $D$
 and is  given by \cite{az,ivanov}
\begin{eqnarray}\label{3.3.mic.2}
\rho(x)=\frac{1}{\pi}+\frac{\sin(2x)}{2\pi x}.
\end{eqnarray}
In Fig.~\ref{fig2}c we compare this analytical result to 
strong coupling lattice simulations for naive quarks in 
the adjoint representation
of $\SU(3)$. The lattice data show excellent agreement for the low-lying Dirac spectrum. Moreover the simulations
 confirm the double degeneracy of the Dirac operator (eigenvalues have also the degeneracy two) and the fact that there are no generic zero modes.

\subsubsection{The Even-Even Case}\label{sec3.3.ee}

Let $L_1$ and $L_2$ be even. This is the case related to the staggered Dirac operator. With help of the symmetries~\eqref{symgenlat3} and \eqref{symmetry-lat-bet4b} the lattice Dirac operator  can, by choosing a particular gauge field independent basis, be brought to the form
\begin{equation}\label{Dirac-lat-bet4-ee}
 D=\diag(U_5^{(1)},U_5^{(1)\,-1})\left(\begin{array}{c|c} 0 & \begin{array}{cc} 0 & \widetilde{W} \\ \widetilde{W}^\dagger & 0 \end{array} \\ \hline \begin{array}{cc} 0 & -\widetilde{W} \\ -\widetilde{W}^\dagger & 0 \end{array} & 0 \end{array}\right)\diag(U_5^{(1)\,-1},U_5^{(1)}),
\end{equation}
 where $\widetilde{W}$ is a real $L_1L_2(N_{\rm c}^2-1)/2\times L_1L_2(N_{\rm c}^2-1)/2$ matrix without any additional restrictions. The additional 
chiral structure is related to the parity of the lattice sites.
 
 The unitary transformation $\diag(U_5^{(1)\,-1},U_5^{(1)})$ does not change the spectrum. Therefore the naive lattice Dirac operator~\eqref{Dirac-lat-bet4-ee} is in the class of chGOE with index $\nu=0$
 (because $\widetilde{W}$ is a square matrix). The Dirac spectrum is doubly degenerate which is taken care of when constructing the staggered Dirac operator. The symmetry breaking pattern is
$\U(4N_{\rm f}) \to \USp(4N_{\rm f} )$ \cite{V} and the microscopic spectral density is given by the $\nu =0 $ result of the chGOE \cite{V2} 
\begin{eqnarray}\label{3.3.mic.3}
\rho(x)=\frac{x}{2}\left[J_{0}^2(x)-J_{1}^2(x)\right]+\frac{1}{2}J_{0}(x)\left[1-\int\limits_0^{x}J_{0}(\widetilde{x})d\widetilde{x}\right],
\end{eqnarray}
 Therefore the level repulsion is linear, the levels have no repulsion from the origin and there are no generic zero modes. The analytical result~\eqref{3.3.mic.3} is compared with lattice data in Fig.~\ref{fig2}d showing a perfect agreement.

\subsection{QCD with more than Two Colors and Fermions in the Fundamental Representation}\label{sec3.4}

In this case there are no anti-unitary symmetries. The  structure and the symmetry class of the Dirac operator are only related to the parity of the lattice. Hence, we have to take the structure of the naive lattice Dirac operator as  shown in Eqs.~\eqref{symgenlat1}, \eqref{symgenlat2}, and \eqref{symgenlat3}.

The odd-odd and even-even lattices are in the same universality class
and are both discussed in subsection~\ref{sec3.4.ooee}. The case of one even 
number of lattice sites and one odd number is considered in subsection~\ref{sec3.4.eo}.

\subsubsection{The Odd-Odd and Even-Even Case}\label{sec3.4.ooee}

If the parity of both directions is odd, there are no additional 
symmetries and we are   
in the  universality class of chGUE with the symmetry breaking pattern
$\U(N_{\rm f})\times \U(N_{\rm f}) \rightarrow \U(N_{\rm f})$. The Dirac operator has the form~\eqref{formgenlat1}. The eigenvalues of $D$  show no degeneracies  and the microscopic spectral
density is given by  the $\nu =0$ result of chGUE \cite{VZ}
\begin{eqnarray}\label{3.4.mic.1}
\rho(x)=\frac{x}{2}\left[J_{0}^2(x)+J_{1}^2(x)\right].
\end{eqnarray}
Note that the two-dimensional Dirac operator has no zero modes.
Therefore the level repulsion is quadratic and the repulsion from the origin is linear.

If both numbers of lattice sites, $L_1$ and $L_2$, are even, the off-diagonal block $W$
becomes itself chiral and the Dirac-operator takes the form~\eqref{formgenlat3}. Since we have no additional symmetries the symmetry class is again the one of chGUE. 
The only difference with  the odd-odd case is 
a doubling of the number of flavors with the chiral symmetry breaking pattern  $\U(2N_{\rm f})\times \U(2N_{\rm f}) \rightarrow \U(2N_{\rm f})$. 
Apart from an additional degeneracy from the doubling of the flavors,
the spectral properties remain the same. In particular, the microscopic spectral density has index $\nu = 0$ and is given by  Eq.~\eqref{3.4.mic.1}.

\begin{figure}[!ht]
\centerline{\includegraphics[width=0.5\textwidth,clip=]{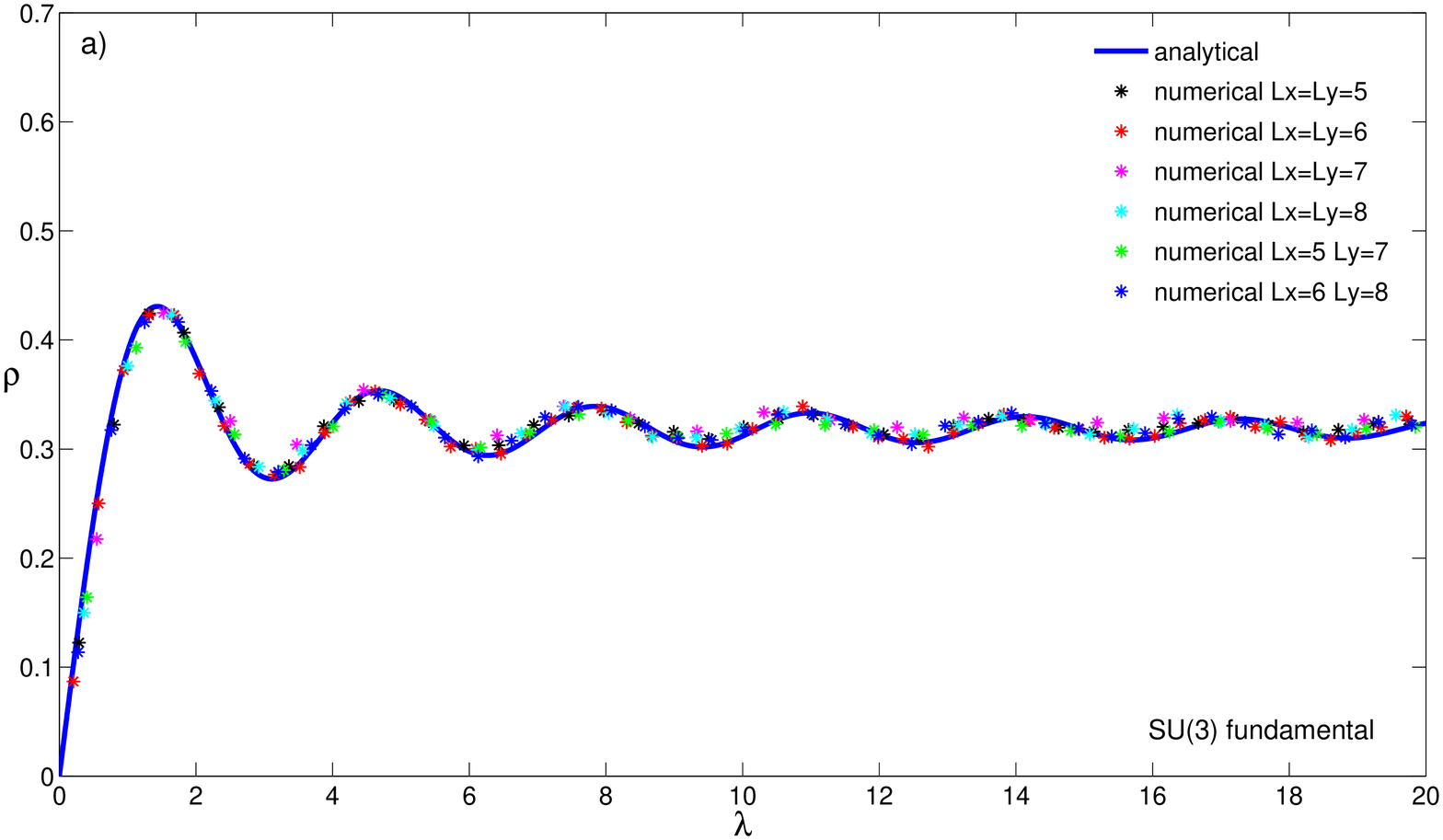}
\includegraphics[width=0.5\textwidth,clip=]{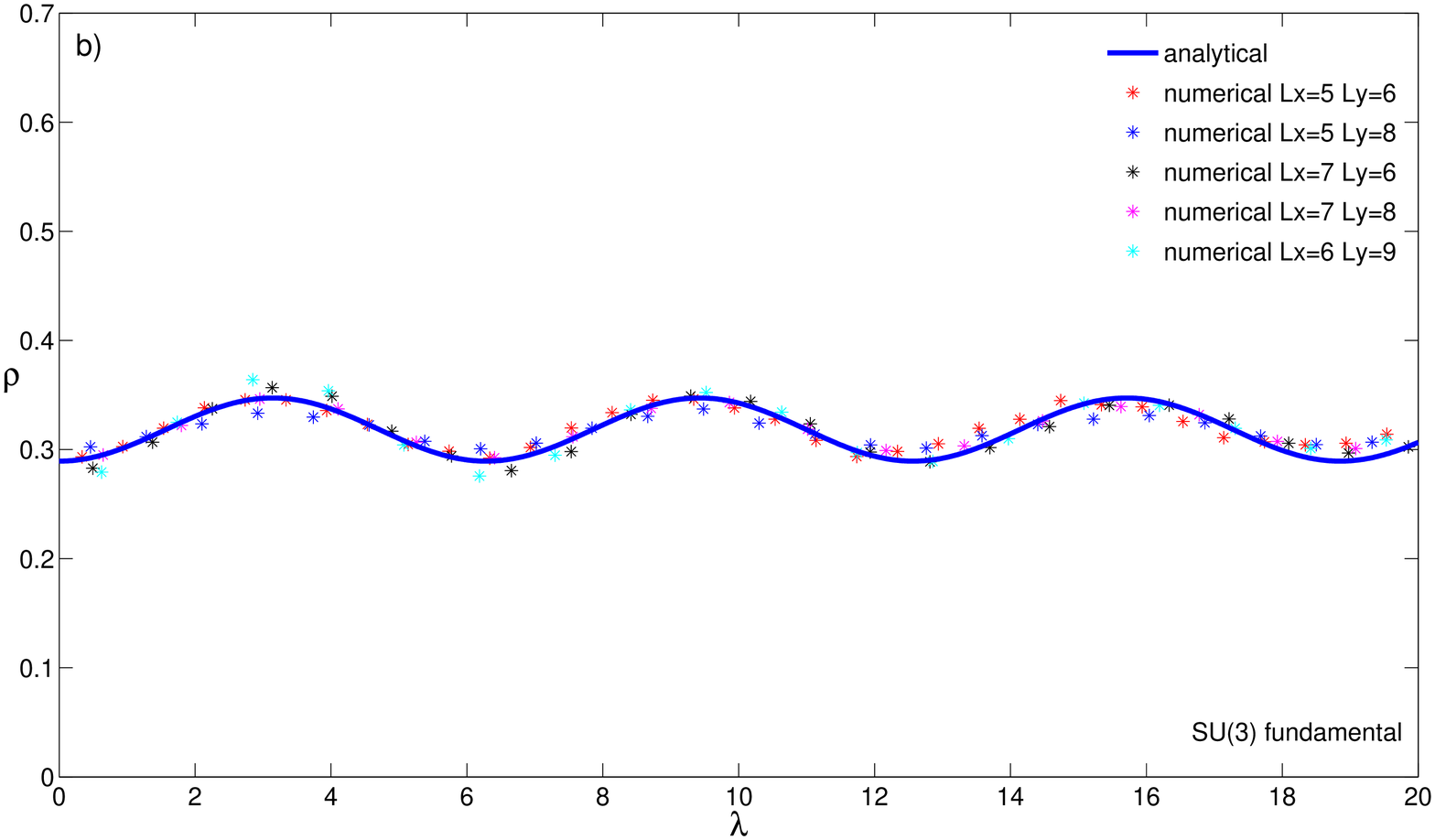}}
\caption{ Comparison of the microscopic level density  of lattice QCD data in the strong coupling limit at various lattice sizes (stars)
 and the analytical results predicted by the corresponding random matrix theories (solid curves). The presented lattice gauge theories are: a) $\SU(3)$ fundamental with $L_1+L_2=$ even and b) $\SU(3)$ fundamental with $L_1+L_2=$ odd. Note that in figure b) we have a strong oscillation on top of 
the universal result which is a constant equal to $1/\pi$. Therefore we plotted the GUE result with its first correction in a $1/n$ expansion in its matrix size $n$. Astoundingly also this non-universal term seems to fit the lattice data quite  well.}
\label{fig3}
\end{figure}

In Fig.~\ref{fig3}a we show lattice data for the spectral density of the Dirac 
operator in the case that both $L_1$ and $ L_2$ are either odd or even. There is an excellent agreement with the
analytical random matrix result~\eqref{3.4.mic.1}. 
Also the degree of degeneracy and the fact that there are  no zero modes is confirmed by the lattice simulations.

\subsubsection{The Even-Odd Case}\label{sec3.4.eo}

The situation changes if $L_1+L_2$ is odd. Then the Dirac operator $D$ follows the structure~\eqref{formgenlat2} where the $L_1L_2N_{\rm c}\times L_1L_2N_{\rm c}$ matrix $H$ is Hermitian. The corresponding symmetry class is represented by the GUE and denoted by the Cartan symbol A \cite{class}. Due to the structure~\eqref{formgenlat2} and the Hermiticity of $H$, the flavor symmetry is doubled to $\U(2N_{\rm f})$. However the eigenvalues of $D$ are not doubly degenerate but come in complex conjugate pairs $\pm\imath\lambda$  because $H$ appears in the off-diagonal blocks.

The lattice Dirac operator is in the same universality class as the three dimensional continuum theory.  Hence the symmetry breaking pattern for this
case is already  known from QCD in three dimensions \cite{VZ3}. A nonzero
 spectral density of $H$ results in a nonzero spectral density of 
the Dirac operator (see the discussion in subsection~\ref{sec3.2.oe})
resulting in a chiral condensate with source term $m$.
 The chiral condensate is invariant under a transformation with the unitary matrix $U\in\U(2N_{\rm f})$ if it fulfills
\be\label{symrelbet2}
U^\dagger\bmat 0 & m\eins_{N_{\rm f}} \\ -m \eins_{N_{\rm f}} & 0 \emat  
U =\bmat 0 & m\eins_{N_{\rm f}} \\ -m \eins_{N_{\rm f}} & 0 \emat.
\ee
This breaks chiral symmetry 
according to the pattern $\U(2N_{\rm f})\rightarrow \U(N_{\rm f})\times \U(N_{\rm f})$.

The microscopic level density including the $O(1/n)$ corrections of a $2n\times 2n$ GUE is given by
\begin{eqnarray}\label{3.4.mic.2}
\rho(x)=\frac{1}{\pi}\left[1+\frac{\cos 2x}{8n}\right].
\end{eqnarray}
In order
to obtain a better fit of the analytical result to the lattice  data, we 
have included the correction term  multiplied by a fitting parameter.
 In Fig.~\ref{fig3}b we compare the microscopic level density of GUE and 
lattice results. The lattice data exhibit much larger oscillations than the ones
given by the  $O(1/n)$ correction in Eq.~\eqref{3.4.mic.2}. One possible mechanism that may contribute to this enhancement
is the condition that the off-diagonal block $H$ of $D$ is traceless, $\tr  H =0 $, since the translation matrices~\eqref{3.1.1} have no diagonal elements. In appendix~\ref{app2} we evaluate the spectral density for
the random matrix ensemble that interpolates between the GUE and the traceless GUE.
The result is given by
\begin{eqnarray}
 \rho_t(x) &=&\frac{1}{\pi}\left(1+\frac{1}{8n}\exp\left[\frac{2t}{t+1}\right]\cos\left[2 x\right]\right),\label{leveldensity-ft}
\end{eqnarray}
which shows oscillations that are  enhanced by a factor of $e^2\approx7.4$ 
for a 
traceless random matrix ($t\to\infty$) in comparison to the original 
GUE ($t=0$). 
Because the lattice Dirac operator is sparse
the effective value of $n$ is expected much less than the size of the matrix. Nevertheless  we would also expect that $n$ still increases with the lattice size. However when using $n$ in Eq.~(\ref{leveldensity-ft}) as a fitting parameter 
we find that $n\approx 7$ for almost all simulations. It is not clear why the amplitude of the
oscillations does not depend on the lattice size which should be analyzed in more detail.
Also other conditions such as the fixed  Euclidean norm of $ H$, i.e. $\tr H^2= 4N_{\rm c}L_1L_2$, may contribute to the amplitude of the oscillations.

\section{Conclusions}

We have analyzed quenched two-dimensional lattice QCD Dirac spectra at strong coupling.
The main differences with QCD in four dimensions are the absence of Goldstone bosons, the absence of topology corresponding to the Atiyah-Singer index theorem, and the non-commutativity of the
anti-unitary symmetries and the axial symmetry. As is the case in four dimensions, the 
symmetries of the Dirac operator depend on the parity  of the number of lattice points 
in each direction. However in two dimensions we find a much richer classification of  symmetry breaking patterns. 
As is the case in four dimensions, the corresponding random matrix class is determined
by the anti-unitary and the involutive symmetries. 
 This is consistent with the
maximum spontaneous breaking of chiral symmetry.

The simulations were performed with periodic boundary conditions in both directions even though we have also checked the effect of anti-periodicity in one direction. Our results remain unaffected in terms of the identifications of the universality class. Only a marginal increase of the Thouless energy was observed by this modification.

Notwithstanding the Mermin-Wagner-Coleman theorem,  
we find that the agreement with  random matrix theory is qualitatively
the same in two and four dimensions.  The agreement is particularly good if the
Goldstone manifold contains a $\U(1)$ or $\Ort(1)\simeq\mathbb{Z}_2$  group 
(i.e for the classes $D$, $DII$, $BDI$, $CII$ and $AIII$). 
This raises the possibility that the long 
range correlations that give rise to random matrix statistics are related 
to the topological properties of the Goldstone manifold \cite{Gruzberg:2012fb}.

In this paper all numerical results are
at nonzero lattice spacing. We did not attempt to perform an extrapolation 
to the continuum limit. 
Based on a bosonized form of the  QCD partition function
in terms of hadronic fields, one would expect a domain of low-lying eigenvalues
that is dominated by the fluctuations of the zero momentum modes so 
that they are
 correlated
according to random matrix theory. In the continuum limit the two dimensional theory is expected
to renormalize to a theory without spontaneous symmetry breaking. What is disturbing is that
we do not observe a qualitative different behavior between QCD in two and four dimensions.

Since quenched spectra are obtained by a supersymmetric extension of the
partition function, our results seem to favor the suggestion by Niedermaier
and Seiler that noncompact symmetries can be broken
 spontaneously  in two dimensions. 
One of the signatures of this type of spontaneous
symmetry breaking is an order parameter that wanders off to infinity. Indeed,
in \cite{damgaard} it was found that the chiral condensate of the quenched Schwinger
model seems to diverge in the thermodynamic limit. On the other hand,
the Dirac spectrum of the $N_{\rm f} =1 $
Schwinger model behaves as predicted by random matrix theory. It is clear
that the chiral condensate is determined by the anomaly and does not
involve any noncompact symmetries. Because of the absence of massless
excitations the partition function of the one flavor Schwinger
model must be smooth as a function of the quark mass. This implies
that the condensate due to the nonzero Dirac eigenvalues must be the
same as the condensate from the one-instanton configurations
in the massless limit. This suggests that the eigenfunctions of the low-lying
nonzero mode states must be delocalized and that the
eigenvalue fluctuations are described by random matrix theory, so that
the supersymmetric partition function that generates the Dirac spectrum
looks like it has spontaneous symmetry breaking.

An alternative scenario arises because of  
the finiteness of the Thouless energy 
 in units of the average level
spacing. 
The fermion determinant due to massless
quarks may push all eigenvalues beyond the Thouless energy   
into the localized domain  resulting in a partition function with no spontaneous breaking of chiral symmetry. To find out if this is the case we would have
to study two-dimensional lattice QCD with dynamical quarks. 
This scenario is not favored by simulations of the
Schwinger model. Both the one- and two-flavor Schwinger model show
excellent agreement with random matrix statistics and the agreement improves
with increasing volumes which also excludes the possibility that the localization
length is larger than the size of the box.

Our study raises many questions. The most fundamental issue is 
the reconciliation of the agreement with random matrix theory and the implied
spontaneous breaking of chiral symmetry with the Mermin-Wagner-Coleman
theorem. In particular, 
can the noncompact symmetry of the supersymmetric generating function for
the Dirac spectrum of two-dimensional QCD-like theories be spontaneously 
broken? To address this we have to analyze the approach to the
thermodynamic limit and the continuum limit. Such studies could also settle whether or not
 the localization length of the low-lying states exceeds the size of
the box used in the present work. This is supported by Dirac spectra of the  
quenched Schwinger model which deviate more from random matrix theory with 
increasing volume \cite{damgaard}, but there is no hint of this in our results. 
Another intriguing question is the possibility that
all states become localized beyond a critical number of flavors.
A final issue concerns the number of generic zero modes of the QCD Dirac operator for fermions in the adjoint representation.
 With chiral perturbation theory and random matrix theory we predict 
that the Dirac operator may have no or only two generic zero modes of opposite chirality. In future work we hope to address the  nature of
these zero modes and the possible relation with the complexified
zero modes found in Ref.~\cite{smilga}.

\section{Acknowledgments}
This work was supported by U.S. DOE Grant No. DE-FG-88ER40388
and the  Humboldt Foundation (MK). Gernot Akemann, Alexander Altland, Poul Damgaard, Erhard Seiler, Andrei Smilga, Peter van Nieuwenhuizen and
Martin Zirnbauer are thanked for stimulating discussions.

\appendix

\section{Random matrix theories of two-dimensional continuum QCD}\label{app1}

In this appendix  we evaluate the joint probability density of the eigenvalues and the pattern of chiral symmetry breaking  
of random matrix theory corresponding to the continuum limit of two dimensional QCD. The case of two colors with fundamental fermions is worked out  in the subsection~\ref{app1.a} and the case with two or more colors with fermions in the
adjoint representation is discussed in the subsection~\ref{app1.b}. The case with three or more colors
with fermions in the fundamental representation follows the same pattern in two and four dimensions
and is not discussed here. Although the results of this appendix are known, discussing them in the present framework
will add to the readability of this paper.

\subsection{Random Matrix Theory for Two-Dimensional QCD with Two Colors in the Fundamental Representation}\label{app1.a}

For two colors with the quarks in the fundamental representation we can find a gauge field independent basis
for which the Dirac operator becomes real. In two dimensions this transformation does not commute with the transformation to a block structure reflecting its chiral symmetry, see section~\ref{sec:cont.1}. We choose to preserve the chiral structure of the Dirac operator. Then
the consequence of the anti-unitary symmetry is that the off-diagonal block of the Dirac operator is complex  anti-self-dual which is unitarily equivalent
to a random matrix theory with an off-diagonal block that is complex symmetric.
The corresponding chiral random matrix theory is given by
\begin{eqnarray}\label{2.1.2}
 D=\left[\begin{array}{cc} 0 & W \\ - W^\dagger & 0 \end{array} \right],\quad W=-\tau_2 W^T\tau_2\in\mathbb{C}^{2n\times2n},
\end{eqnarray}
or equivalently by
\begin{eqnarray}\label{2.1.2b}
 D'=\left[\begin{array}{cc} 0 & W \tau_2 \\ -W^\dagger\tau_2 & 0 \end{array} \right],\quad (W\tau_2)^T= W\tau_2  \in\mathbb{C}^{2n\times2n}.
\end{eqnarray}
The probability distribution is taken to be Gaussian
\begin{eqnarray}\label{2.1.1}
 P(W)d[W]\propto\exp\left[-n\tr  WW^\dagger\right]\prod\limits_{1\leq i\leq j\leq 2n}d\,\RE\,W_{ij}d\,\IM\,W_{ij}.
\end{eqnarray}
In the subsection~\ref{app1.a.a} we calculate the joint eigenvalue probability density of this theory 
(see Ref.~\cite{az}).  In 
the subsection~\ref{app1.a.b} we rederive its partition function which was already summarized for all chiral ensembles in Ref.~\cite{class,ivanov}.

\subsubsection{ Joint Probability Density}\label{app1.a.a}

The joint probability density of the eigenvalues of the random matrix $D$ denoted by $p(\Lambda)$ is defined by
\begin{eqnarray}\label{2.1.3}
 \int\limits_{\mathbb{C}^{2n\times 2n}} f(D) P(W)d[W]=\int\limits_{\mathbb{R}_+^{2n}} f(\pm\imath\Lambda) p(\Lambda)\prod\limits_{1\leq j\leq 2n}d\lambda_j
\end{eqnarray}
for any function $f$ invariant under
\begin{eqnarray}\label{2.1.3.b}
 f(D)=f(VDV^\dagger)
\end{eqnarray}
for all $V=\diag(\widetilde{V},\tau_2\widetilde{V}^*\tau_2)$ 
or $W \rightarrow \widetilde{V} W\tau_2 \widetilde{V}^T\tau_2$
with $\widetilde{V}\in\U(2n)$.

The characteristic polynomial of $D$ can be rewritten as
\begin{eqnarray}\label{2.1.4}
 \det(D-\imath\lambda\eins_{4n})=\det(WW^\dagger-\lambda^2\eins_{2n})=\det(W^\dagger W-\lambda^2\eins_{2n}).
\end{eqnarray}
Let $U\in\U(2n)/\U^{2n}(1)$ be the matrix diagonalizing 
$WW^\dagger$, i.e. $WW^\dagger=U\Lambda^2 U^\dagger$ with the 
positive definite, diagonal matrix $\Lambda^2\in\mathbb{R}_+^{2n}$. Then we can relate the eigenvectors of $WW^\dagger$ to those of $W^\dagger W$. Let
\be
\label{2.1.5}
 WW^\dagger U=(W\tau_2)(W\tau_2)^{\dagger}=U\Lambda^2,
\ee
 then  complex conjugation results in
\be
 (W\tau_2)^*(W\tau_2)^T U^*=U^*\Lambda^2, 
\ee
and because of the symmetry of $W\tau_2$, we also have
\be
 (W\tau_2)^\dagger W\tau_2 U^*=U^*\Lambda^2.
\ee
Hence the eigenvalue decomposition of $W^\dagger W$ reads
\begin{eqnarray}\label{2.1.6}
 (W\tau_2)^\dagger (W\tau_2) = U^*\Lambda^2 U^T.
\end{eqnarray}
The combination of this decomposition with $ WW^\dagger =U\Lambda^2U^\dagger$
yields a singular value decomposition of $W$,
\begin{eqnarray}\label{2.1.7}
  W\tau_2 &=&UZ U^T
\end{eqnarray}
with the complex, diagonal matrix $Z\in\mathbb{C}^{2n}$ such that $|Z|=\Lambda$ and $U\in\U(2n)/\U^{2n}(1)$. The number of
degrees of freedom is $2n(2n+1)$ on both sides of Eq.~\eqref{2.1.7}. Hence, the right hand side of Eq.~\eqref{2.1.7} can be used as a parameterization of $W$. The phases of $Z$ can be absorbed in $U$ so that $W$ can be parameterized as
\begin{eqnarray}\label{2.1.8}
  W\tau_2 &=&U\Lambda U^T
\end{eqnarray}
with the positive definite, diagonal matrix $\Lambda\in\mathbb{R}_+^{2n}$ and $U\in\U(2n)$.

In the next step we calculate the invariant length element which directly yields the Haar measure of $W$ in the coordinates~\eqref{2.1.8},
\begin{eqnarray}\label{2.1.9}
  \tr  dWdW^\dagger &=&\tr  d(W\tau_2)d(W\tau_2)^\dagger \\
  &=&\tr  d\Lambda^2+\tr  \left(U^\dagger dU \Lambda+\Lambda (U^\dagger dU)^T\right)\left(U^\dagger dU \Lambda+\Lambda (U^\dagger dU)^T\right)^\dagger\nonumber\\
   &=&\sum\limits_{1\leq i\leq 2n} (d\lambda_i^2+4\lambda_i^2(U^\dagger dU)_{ii}^2)\nonumber\\
  &&\hspace*{-1cm}+\sum\limits_{1\leq i< j\leq 2n}\left[\begin{array}{cc} (U^\dagger dU)_{ij}, & (U^\dagger dU)_{ij}^* \end{array}\right]\left[\begin{array}{cc} \lambda_i\lambda_j & \displaystyle-\frac{\lambda_i^2+\lambda_j^2}{2}\\ \displaystyle-\frac{\lambda_i^2+\lambda_j^2}{2} & \lambda_i\lambda_j \end{array}\right]\left[\begin{array}{c} (U^\dagger dU)_{ij}, \\ (U^\dagger dU)_{ij}^* \end{array}\right]\nonumber.
\end{eqnarray}
Note that the Pauli matrix $\tau_2$ drops out. Moreover we have used the anti-Hermiticity of $U^\dagger dU$. From the invariant length~\eqref{2.1.9} we find the joint probability density
\begin{eqnarray}\label{2.1.10}
 p(\Lambda)\prod\limits_{1\leq j\leq 2n}d\lambda_j\propto|\Delta_{2n}(\Lambda^2)|\prod\limits_{1\leq j\leq 2n}\exp\left[-n\lambda_j^2\right]\lambda_jd\lambda_j,
\end{eqnarray}
cf. Ref.~\cite{az,ivanov}. This coincides with  the joint probability density of the nonzero eigenvalues of the
 chiral GOE with $\nu=1$, which has one zero mode while the present model has no zero modes at all. 
Its microscopic spectral 
density has a linear slope at the origin and the level repulsion 
is also linear at small distances, cf. Fig~\ref{fig1}a.

\subsubsection{Partition Function}\label{app1.a.b}

The partition function  with $N_{\rm f}$ flavors is defined by
\be\label{partfunc1}
Z(N_{\rm f}) = \int d[W] \prod_{k=1}^{N_{\rm f}} \det(D+m_k\eins_{4n}) P(W).
\ee
Due to the decomposition~\eqref{2.1.8} we multiply $D$ by the unitary matrix $\diag(\eins_{2n},\tau_2)$ from the left and from the right which keeps the spectrum invariant. To evaluate the average~\eqref{partfunc1}
 we first rewrite the determinants as Gaussians over Grassmann variables
\begin{eqnarray}
  Z(M)&\propto&\int d[W,V] \exp\left[-n\tr  W\tau_2(W\tau_2)^\dagger\right]\label{2.1.12}\\
  &&\times\exp\left[\tr  V_R^\dagger W\tau_2V_L-\tr  V_L^\dagger (W\tau_2)^\dagger V_R+\tr  M(V_R^\dagger V_R+V_L^\dagger V_L)\right]\nonumber
\end{eqnarray}
 with the mass matrix $M=\diag(m_1,\ldots,m_{N_{\rm f}})$. The matrices $V_R$ and $V_L$ are both $2n\times N_{\rm f}$ rectangular matrices comprising independent Grassmann variables as matrix elements. Because $W\tau_2$ is symmetric we have to symmetrize the matrices $V_LV_R^\dagger$ and $V_RV_L^\dagger$. After  integrating over $W$ we obtain
\begin{eqnarray}
  Z(M)&\propto&\int d[V]\exp\left[-\frac{1}{4n}\tr  (V_LV_R^\dagger-V_R^*V_L^T)(V_RV_L^\dagger-V_L^*V_R^T)+\tr  M(V_R^\dagger V_R+V_L^\dagger V_L)\right]\nonumber\\
 &\propto&\int d[V] \exp\left[\frac{1}{4n}\tr  (\widetilde{\tau}_2\otimes\eins_{N_{\rm f}})\sigma(\widetilde{\tau}_2\otimes\eins_{N_{\rm f}})\sigma^T+\tr  (\eins_2\otimes M) \sigma\right],\label{2.1.13}
\end{eqnarray}
where $\tau_2$ completely drops out. The second Pauli matrix $\widetilde{\tau}_2$ acts on flavor space and should not be confused with $\tau_2$ which acts on color space for QCD and its analogue in random matrix theory. The dyadic super matrix
\begin{eqnarray}
 \sigma=\left[\begin{array}{c} V_R^\dagger \\ -V_L^T\end{array}\right]\left[\begin{array}{cc} V_R, & V_L^*\end{array}\right].\label{2.1.14}
\end{eqnarray}
is nilpotent and can be replaced by a unitary matrix $U\in\U(2N_{\rm f})$ via the superbosonization formula \cite{supersom,superzirn,supermario}. 
By rescaling $U\to 2n U$ and introducing the rescaled mass matrix $\widehat{M}=2n M$, we arrive at
\begin{eqnarray}
  Z(\widehat{M})&\propto&\int\limits_{\U(2N_{\rm f})} \exp\left[n\tr  (\widetilde{\tau}_2\otimes\eins_{N_{\rm f}})
U(\widetilde{\tau}_2\otimes\eins_{N_{\rm f}})U^T+\tr  (\eins_2\otimes\widehat{M})U\right]{\det}^{-2n}U d\mu(U),\hspace*{0.5cm}\label{2.1.15}
\end{eqnarray}
where $d\mu$ is the normalized Haar-measure.

In the microscopic limit ($n\to\infty$ and $\widehat{M}$ fixed) we can apply the saddlepoint approximation. The saddlepoint equation is given by
\begin{eqnarray}
 U^{-1}=(\widetilde{\tau}_2\otimes\eins_{N_{\rm f}})U^T (\widetilde{\tau}_2\otimes\eins_{N_{\rm f}}).\label{2.1.16}
\end{eqnarray}
Since $U\in\U(2N_{\rm f})$ Eq.~\eqref{2.1.16} implies $U\in\USp(2N_{\rm f})$. The final result is given by
\begin{eqnarray}
 Z(\widehat{M})&=&\int\limits_{\USp(2N_{\rm f})} \exp\left[\tr  (\eins_2\otimes\widehat{M})U\right]d\mu(U)\label{2.1.17}\\
&=&\int\limits_{\USp(2N_{\rm f})} \exp\left[\frac{1}{2}\tr  (\eins_2\otimes\widehat{M})(U+U^{-1})\right]d\mu(U).\nonumber
\end{eqnarray}
Although the joint probability density of the eigenvalues coincides with chGOE, the chiral 
symmetry breaking pattern ($\USp(2N_{\rm f})\times\USp(2N_{\rm f})\to\USp(2N_{\rm f})$) turns out to be different and agrees with Ref.~\cite{class,imbo}. Especially there are no zero modes such that the
partition function does not vanish at $M=0$ which would be the case for chGOE with the index $\nu=1$, see Ref.~\cite{V}.

\subsection{Two Dimensional QCD in the Adjoint Representation}\label{app1.b}

For two dimensional QCD with quarks in the adjoint representation the anti-unitary symmetry of the
Dirac operator allows us to choose a gauge field independent 
basis for which the Dirac operator becomes quaternion real.
However, when performing this transformation we will lose the chiral block structure. We choose to preserve this structure. Then
the anti-unitary symmetry requires that the off-diagonal block of the Dirac operator becomes anti-symmetric.
The corresponding random matrix theory is given by
\begin{eqnarray}\label{2.2.2}
 D=\left[\begin{array}{cc} 0 & W \\ -W^\dagger & 0 \end{array} \right],\quad W=-W^T\in\mathbb{C}^{(2n+\nu)\times (2n+\nu)}.
\end{eqnarray}
with the probability distribution
\begin{eqnarray}\label{2.2.1}
 P(W)d[W]\propto\exp\left[-n\tr  WW^\dagger\right]\prod\limits_{1\leq i< j\leq (2n+\nu)}d\,\RE\,W_{ij}d\,\IM\,W_{ij}.
\end{eqnarray}
Because odd-dimensional anti-symmetric matrices have one generic zero eigenvalue we have to
distinguish the even and odd dimensional case (denoted by $\nu=0$ and $\nu=1$ respectively).

In subsection~\ref{app1.b.a} we evaluate the joint probability density of the eigenvalues and in subsection~\ref{app1.b.b} we
discuss the corresponding partition function for $\nu =0,1$. 
These results were obtained previously in Refs.~\cite{class,az,ivanov,imbo}. 

\subsubsection{Joint Probability Distribution}\label{app1.b.a}

The joint probability density $p(\Lambda)$ is defined as in Eq.~\eqref{2.1.3} while 
 the arbitrary function $f$ has the invariance
\begin{eqnarray}\label{2.1.3.c}
 f(D)=f(VDV^\dagger),\quad \forall\, V=\diag(\widetilde{V},\widetilde{V}^*)\ {\rm with}\ \widetilde{V}\in\U(2n+\nu).
\end{eqnarray}

Let $\nu=0$, i.e. $W$ is even dimensional. Analogous to the discussion in subsection~\ref{app1.a.a} we can quasi-diagonalize $W$, i.e.
\begin{eqnarray}\label{2.2.3}
  W &=&U(\tau_2\otimes\Lambda )U^T
\end{eqnarray}
with a positive definite, diagonal matrix $\Lambda\in\mathbb{R}_+^{n}$ and 
 unitary matrix $U\in\U(2n)/{\rm SU}^n(2)$. 
The division with the subgroup ${\rm SU}^n(2)$ is the result of 
the identity $\widetilde{U}\tau_2\widetilde{U}^T=\tau_2$ for all $\widetilde{U}\in{\rm SU}(2)$.

The matrix $\tau_2\otimes\Lambda$ has $\pm\lambda_j$ as eigenvalues. 
We can  use the result~\eqref{2.1.10} by replacing $\diag(\lambda_1,\ldots,\lambda_{2n})\to\diag(\lambda_1,\ldots,\lambda_{n},-\lambda_1,\ldots,-\lambda_n)$ and taking care of the fact that some degrees of freedom of $\U(2n)$ are missing. We can apply the result~\eqref{2.1.10} because the 
invariant length element is calculated similar to the $(\beta_{\rm D}=1,\ d=2)$-case. 
Hence, we find the joint probability density 
\begin{eqnarray}\label{2.2.4}
 p(\Lambda)\prod\limits_{1\leq j\leq 2n}d\lambda_j\propto\Delta_{n}^4(\Lambda^2)\prod\limits_{1\leq j\leq n}\exp\left[-n\lambda_j^2\right]\lambda_jd\lambda_j,
\end{eqnarray}
cf. Ref.~\cite{az,ivanov}. This density coincides with the chGSE for $\nu=-1/2$. 

Let us consider the case with an odd dimension, $W=-W^T\in\mathbb{C}^{(2n+1)\times(2n+1)}$. Since an odd dimensional anti-symmetric matrix has one generic zero mode we have to modify the decomposition~\eqref{2.2.3} according to
\begin{eqnarray}\label{2.2.3.b}
  W &=&U\diag(\tau_2\otimes\Lambda,0 )U^T,
\end{eqnarray}
where $\Lambda\in\mathbb{R}_+^{n}$ and $U\in\U(2n+1)/[{\rm SU}^n(2)\times\U(1)]$. Hence, the joint probability density~\eqref{2.2.4} becomes \cite{az,ivanov}
\begin{eqnarray}\label{2.3.1}
 p(\Lambda)\prod\limits_{1\leq j\leq 2n}d\lambda_j\propto\Delta_{n}^4(\Lambda^2)\prod\limits_{1\leq j\leq n}\exp\left[-n\lambda_j^2\right]\lambda_j^5d\lambda_j
\end{eqnarray}
by employing the result~\eqref{2.1.10}  with the replacement $\diag(\lambda_1,\ldots,\lambda_{2n})\to\diag(\lambda_1,\ldots,\lambda_{n},-\lambda_1,\ldots,-\lambda_n,0)$ and taking care of the subgroup ${\rm SU}^n(2)\times\U(1)$ that is
divided out. This coincides with the joint probability density of the non-zero eigenvalues
of  chGSE with $\nu=1/2$.

\subsubsection{Partition Function}\label{app1.b.b}

The partition function with $N_{\rm f}$ fermionic flavors~\eqref{partfunc1}
can be again mapped to flavor space via the rectangular $(2n+\nu)\times N_{\rm f}$ matrices $V_R$ and $V_L$ comprising Grassmann variables only. The analogue of Eq.~\eqref{2.1.13} is given by
\begin{eqnarray}
  Z(M)&\propto&\int d[V]\exp\left[-\frac{1}{4n}\tr  (V_LV_R^\dagger+V_R^*V_L^T)(V_RV_L^\dagger+V_L^*V_R^T)+\tr  M(V_R^\dagger V_R+V_L^\dagger V_L)\right]\nonumber\\
 &\propto&\int \exp\left[\frac{1}{4n}\tr  \sigma\sigma^T+\tr  (\widetilde{\tau}_1\otimes M) \sigma\right]d[V]\label{2.2.6}
\end{eqnarray}
with the dyadic supermatrix
\begin{eqnarray}
 \sigma=\left[\begin{array}{c} -V_L^T \\ V_R^\dagger \end{array}\right]\left[\begin{array}{cc} V_R, & V_L^*\end{array}\right].\label{2.2.7}
\end{eqnarray}
The first Pauli matrix $\widetilde{\tau}_1$ acts on flavor space. The superbosonization formula \cite{supersom,superzirn,supermario} yields
\begin{eqnarray}
  Z(\widehat{M})&\propto&\int\limits_{\U(2N_{\rm f})} \exp\left[n\tr  UU^T+\tr  (\widetilde\tau_1\otimes\widehat{M})U\right]{\det}^{-2n-\nu}U d\mu(U).\label{2.2.8}
\end{eqnarray}
In  the microscopic limit by taking $n$ to infinity we have to solve the saddlepoint equation
\begin{eqnarray}
 U^{-1}=U^T.\label{2.2.9}
\end{eqnarray}
Therefore we end up with an integral over the group ${\rm O}(2N_{\rm f})$, i.e.
\begin{eqnarray}
 Z(\widehat{M})&=&\int\limits_{{\rm O}(2N_{\rm f})} \exp\left[\tr  (\widetilde\tau_1\otimes\widehat{M})U\right]{\det}^\nu Ud\mu(U),\label{2.2.10}\\
&=&\int\limits_{{\rm O}(2N_{\rm f})} \exp\left[\tr  (\eins_2\otimes\widehat{M})U\right]{\det}^\nu Ud\mu(U),\nonumber\\
&=&\int\limits_{{\rm O}(2N_{\rm f})} \exp\left[\frac{1}{2}\tr  (\eins_2\otimes\widehat{M})(U+U^{-1})\right]{\det}^\nu Ud\mu(U)\nonumber
\end{eqnarray}
with $\widehat M = 2nM$. 
For $\nu = 0$ the partition function is of order one for $\widehat{M}\ll1$ 
while for $\nu =1$, the sum over two disconnected components of $\Ort(2N_{\rm f})$,
results in a partition function $Z(\widehat{M})\propto\widehat{M}^2$ for $\widehat M \ll 1$. 
 This property as well as the symmetry breaking pattern ${\rm O}(2N_{\rm f})\times{\rm O}(2N_{\rm f})\to{\rm O}(2N_{\rm f})$ underlines the difference of the random matrix ensemble~\eqref{2.2.2} with chGSE, see Refs.~\cite{V,class}.
The sum over $\nu=0$ and $ \nu =1$ gives an integral over $\SO(2N_{\rm f})$ corresponding
to the symmetry breaking pattern of the full partition function \cite{imbo}.

As was shown in Ref.~\cite{smilga}  gauge fields with nonzero topology exist for two-dimensional QCD with adjoint fermions and both partition functions
for $\nu = 0 $  and $ \nu =1$ are realized. The argument
of Ref.~\cite{smilga} predicting additional values of $\nu$ for $N_{\rm c} >2$ 
seems to be in conflict with chiral perturbation theory \cite{imbo}
and random matrix theory, but we hope to address this puzzle in future work.
In lattice QCD at strong coupling in the case of an odd-odd lattice
only $ \nu =0$ and $\nu =1$ are realized for $N_{\rm c}$ odd and $N_{\rm c}$ even, respectively. Our simulations confirm this prediction, see Fig.~\ref{fig2}.

\section{Corrections to the Traceless Ensemble}\label{app2}

In this appendix we calculate the eigenvalue density including $1/n$ corrections for an even-dimensional GUE
with the additional condition that the trace of the matrices may vanish. This condition is implemented
via a Lagrange multiplier. 
The level density is thus given by the random matrix integral,
\begin{eqnarray}
 \rho_{t}^{(n)}(x)&=&\frac{\int_{\Herm(2n)}d[H] \exp\left[-\tr  H^2/(4n)-t\tr^2  H/(8n^2)\right]\tr\delta(H-x\eins_{2n})}{\int_{\Herm(2n)}d[H] \exp\left[-\tr  H^2/(4n)-t\tr^2  H/(8n^2)\right]}.\label{levdenmod1}
\end{eqnarray}
The parameter $t$ interpolates between the traceless condition ($t\to\infty$) and the ordinary GUE
 ($t\to0$). The square of the trace in $H$ can be linearized by a Gaussian integral over an auxiliary scalar variable $\lambda$, meaning that we can trace back the whole problem to ordinary GUE
\begin{eqnarray}
 \rho_{t}^{(n)}(x)&=&\int\limits_{-\infty}^\infty \sqrt{\frac{1+t}{2 t\pi}}d\lambda \exp\left[-\frac{1+t}{2 t}\lambda^2\right]\rho_{0}^{(n)}(x+\imath\lambda).\label{levdenmod2}
\end{eqnarray}
The level density of GUE is given in terms of Hermite polynomials,
 $H_j(x)=x^j+\ldots$, in the following formula \cite{Mehtabook}
\begin{eqnarray}
 \rho_{0}^{(n)}(x)&=&\frac{(2n)!}{\sqrt{4\pi n}}\exp\left[-\frac{x^2}{4n}\right]\label{levdenmod3}\\
 &&\times\left(\frac{(2n)^{2n-1}}{((2n-1)!)^2}H_{2n-1}^2\left(\frac{x}{\sqrt{2n}}\right)-\frac{(2n)^{2n-1}}{(2n)!(2n-2)!}H_{2n}\left(\frac{x}{\sqrt{2n}}\right)H_{2n-2}\left(\frac{x}{\sqrt{2n}}\right)\right).\nonumber
\end{eqnarray}
The large $n$ asymptotics of $H_{2n}(x/\sqrt{2n})$ where $x$  is fixed can be obtained by  the relation between Hermite polynomials with an even order and the associated Laguerre polynomials, $L_n^{(-1/2)}(x)=x^n+\ldots$,
\begin{eqnarray}\label{relHermLag}
H_{2n}\left(\frac{x}{\sqrt{2n}}\right)&=&2^n L_n^{(-1/2)}\left(\frac{x^2}{2n}\right).
\end{eqnarray}
Note that we employ for both polynomials the monic normalization. The associated Laguerre polynomials $L_n^{(\nu)}$ with a positive integer index $\nu$ have a simple representation as a contour integral,
\begin{eqnarray}\label{contLag}
L_n^{(\nu)}\left(\frac{x^2}{2n}\right)&=& \frac{n!}{(2n)^{n}}\int\limits_0^{2\pi}\frac{d\varphi}{2\pi} e^{\imath\nu\varphi}\left(1-\frac{e^{-\imath\varphi}}{2n}\right)^{n+\nu}\exp[x^2 e^{\imath\varphi}],
\end{eqnarray}
which can be expanded asymptotically
\begin{eqnarray}\label{asympLag}
L_n^{(\nu)}\left(\frac{x^2}{2n}\right)&\overset{n\gg1}{\approx}& \frac{1}{(2n)^{n}}\int\limits_0^{2\pi}\frac{d\varphi}{2\pi} e^{\imath\nu\varphi}\exp\left[x^2 e^{\imath\varphi}-(n+\nu)\left(\frac{e^{-\imath\varphi}}{2n}+\frac{e^{-2\imath\varphi}}{8n^2}\right)\right]\\
&\approx&\frac{1 }{(2n)^{n}}\left(\frac{ J_{-\nu}(x)}{x^\nu}-\frac{\nu J_{1-\nu}(x)}{2n x^{\nu-1}}-\frac{ J_{2-\nu}(x)}{8n x^{\nu-2}}\right).\nonumber
\end{eqnarray}
The functions $J_j$ are the Bessel functions of the first kind and can be analytically continued in their index $j$. For $\nu=-1/2$ the expansion  for the Hermite polynomials reads
\begin{eqnarray}
\frac{1}{n!}H_{2n}\left(\frac{x}{\sqrt{2n}}\right)&\overset{n\gg1}{\approx}&\frac{1 }{n^{n}}\left(\sqrt{x} J_{1/2}(x)+\frac{ x^{3/2}J_{3/2}(x)}{4n }-\frac{x^{5/2} J_{5/2}(x)}{8n }\right)\label{asympHerm}\\
&=&\frac{1 }{n^{n}}\sqrt{\frac{2}{\pi}}\left( \sin x+\frac{1 }{8n }\left((x^2-1)\sin x+x\cos x\right)\right).\nonumber
\end{eqnarray}
From this asymptotic expansion it also follows
\begin{eqnarray}
\frac{1}{(n-1)!}H_{2n-1}\left(\frac{x}{\sqrt{2n}}\right)&=&\sqrt{\frac{n}{2}}\frac{\partial}{\partial x}\frac{1}{n!}H_{2n}\left(\frac{x}{\sqrt{2n}}\right)\label{asympHerm.a}\\
&\overset{n\gg1}{\approx}&\frac{1 }{n^{n}}\sqrt{\frac{n}{\pi}}\left( \cos x+\frac{1 }{8n }\left(x\sin x+x^2\cos x\right)\right).\nonumber
\end{eqnarray}
and
\begin{eqnarray}
\frac{1}{(n-2)!}H_{2n-2}\left(\frac{x}{\sqrt{2n}}\right)&=&\frac{\sqrt{2n}(n-1)}{2n-1}\frac{\partial}{\partial x}\frac{1}{(n-1)!}H_{2n-1}\left(\frac{x}{\sqrt{2n}}\right)\label{asympHerm.b}\\
&\overset{n\gg1}{\approx}&\frac{n-1 }{(2n-1)n^{n-1}}\sqrt{\frac{2}{\pi}}\left( -\sin x+\frac{1 }{8n }\left((1-x^2)\sin x+3x\cos x\right)\right),\nonumber
\end{eqnarray}
with help of the recurrence relation of the Hermite polynomials and 
 the Stirling formula including subleading corrections
\begin{equation}\label{Stirling}
 n!\overset{n\gg1}{\approx}\sqrt{2\pi n} n^n e^{-n}\left(1+\frac{1}{12n}\right).
\end{equation}

Summarizing all these asymptotic expansions and plugging everything into the level density~\eqref{levdenmod3} we find the first correction to the GUE asymptotics
\begin{eqnarray}\label{levdenmod4}
 \rho_{0}^{(n)}(x)&\overset{n\gg1}{\approx}&\frac{1}{\pi}\left(1-\frac{x^2}{4n}\right)\left(1+\frac{1}{8n}\right)\left(\cos^2 x+\frac{1 }{4n }\cos x\left(x\sin x+x^2\cos x\right)\right.\\
 &&\left.-\left( -\sin^2 x+\frac{1 }{4n }\sin x\left((1-x^2)\sin x+x\cos x\right)\right)\right)\nonumber\\
 &\approx&\frac{1}{\pi}\left(1+\frac{\cos 2x}{8n}\right).\nonumber
\end{eqnarray}

One can now perform the integral~\eqref{levdenmod2} which yields the result~\eqref{leveldensity-ft}.


\begin{thebibliography}{11}

\bibitem{GL}
J. Gasser and H.~Leutwyler, Phys. Lett. B {\bf 188}, 477 (1987).


\bibitem{LS}
H.~Leutwyler and A.~Smilga, Phys. Rev. D {\bf 46},  5607 (1992).


\bibitem{VPLB}
 J.~J.~M.~Verbaarschot,
  Phys.\ Lett.\  B {\bf 368}, 137 (1996)
  [arXiv:hep-ph/9509369].

\bibitem{BC}
T. Banks and A. Casher, 
 Nucl. Phys. {\bf B 169}, 103 (1980).

\bibitem{SV}
E.~V.~Shuryak and J.~J.~M.~Verbaarschot,
  Nucl.\ Phys.\  A {\bf 560}, 306 (1993)
  [arXiv:hep-th/9212088].

\bibitem{markos}
P.Marko\v{s} and L. Schweitzer, Phys. B {\bf 407}, 4016 (2012) 
[arXiv:1208.3934].

\bibitem{evangelou}
S.N. Evangelou, Phys. Rev. Lett. {\bf 75}, 2550 (1995).

\bibitem{furusaki}
A. Furusaki, Phys. Rev. Lett. {\bf 82}, 604 (1999) 
[arXiv:cond-mat/9808059].

\bibitem{asada}
  Y.~Asada, K.~Slevin, and T.~Ohtsuki,
  Phys.\ Rev.\ Lett. {\bf 89}, 256601 (2002)
[arXiv: cond-mat/0204544].



 \bibitem{stone}
 A. McKane and M. Stone, Ann. Phys. 131, 36 (1981).

\bibitem{jv-1997}
  J.~J.~M.~Verbaarschot,
  In *Zakopane 1997, New developments in quantum field theory* 187-216
  [arXiv:hep-th/9709032].

\bibitem{seiler}
  A.~Duncan, M.~Niedermaier, and E.~Seiler,
  Nucl.\ Phys.\ B {\bf 720}, 235 (2005)
  [Erratum-ibid.\ B {\bf 758}, 330 (2006)]
  [arXiv:hep-th/0405163].

\bibitem{seiler2} 
  M.~Niedermaier and E.~Seiler,
  Commun.\ Math.\ Phys.\  {\bf 270}, 373 (2007)
  [arXiv:math-ph/0601049].

\bibitem{witten-thir}
  E.~Witten,
  Nucl.\ Phys.\ B {\bf 145}, 110 (1978).

\bibitem{farchioni} 
  F.~Farchioni, I.~Hip, C.~B.~Lang, and M.~Wohlgenannt,
  Nucl.\ Phys.\ B {\bf 549}, 364 (1999)
  [arXiv:hep-lat/9812018].

\bibitem{damgaard}
  P.~H.~Damgaard, U.~M.~Heller, R.~Narayanan, and B.~Svetitsky,
  Phys.\ Rev.\ D {\bf 71}, 114503 (2005)
  [arXiv:hep-lat/0504012].

\bibitem{leonid} 
  L.~Shifrin and J.~J.~M.~Verbaarschot,
  Phys.\ Rev.\ D {\bf 73}, 074008 (2006)
  [arXiv:hep-th/0507220].

\bibitem{bietenholtz} 
  W.~Bietenholz, I.~Hip, S.~Shcheredin, and J.~Volkholz,
  Eur.\ Phys.\ J.\ C {\bf 72}, 1938 (2012)
  [arXiv:1109.2649 [hep-lat]].

\bibitem{smilga1} 
  A.~V.~Smilga,
  Phys.\ Lett.\ B {\bf 278}, 371 (1992).

\bibitem{smil2}
  A.~V.~Smilga,
  Phys.\ Rev.\  D {\bf 46}, 5598 (1992).

\bibitem{Christiansen:1998ax} 
  H.~R.~Christiansen,
  Int.\ J.\ Mod.\ Phys.\ A {\bf 14}, 1379 (1999)
  [arXiv:hep-th/9806219].



\bibitem{Christiansen:1997mu} 
  H.~R.~Christiansen,
  (1997) hep-th/9704020.





\bibitem{Gross:1994mr} 
  D.~J.~Gross and A.~Matytsin,
  Nucl.\ Phys.\ B {\bf 429}, 50 (1994)
  [arXiv:hep-th/9404004].

\bibitem{witten2d} 
E. Witten, J.  Geom. Phys. {\bf 9}, 303 (1992).
 
\bibitem{Burda}
 P.~Bialas, Z.~Burda and B.~Petersson,
  Phys.\ Rev.\ D {\bf 83}, 014507 (2011)
  [arXiv:1006.0360 [hep-lat]].
 
\bibitem{BKPW}
   F.~Bruckmann, S.~Keppeler, M.~Panero and T.~Wettig,
  Phys.\ Rev.\ D {\bf 78}, 034503 (2008)
  [arXiv:0804.3929 [hep-lat]].


\bibitem{simons-altland} 
  A.~Altland and B.~D.~Simons,
  Nucl.\ Phys.\ B {\bf 562}, 445 (1999)
  [cond-mat/9909152].

\bibitem{zirn-cf}M.R. Zirnbauer, J.Phys. {\bf A29}, 19 (1996).

\bibitem{wettig-wei} 
  Y.~Wei and T.~Wettig,
  J.\ Math.\ Phys.\  {\bf 46}, 072306 (2005)
  [hep-lat/0411038].

\bibitem{wettig-sl1} 
  B.~Schlittgen and T.~Wettig,
  Nucl.\ Phys.\ B {\bf 632}, 155 (2002)
  [hep-lat/0111039].


\bibitem{imbo} 
  R.~DeJonghe, K.~Frey and T.~Imbo,
  Phys.\ Lett.\ B {\bf 718}, 603 (2012)
  [arXiv:1207.6547 [hep-th]].

\bibitem{ludwig}
A. P. Schnyder, S. Ryu, A. Furusaki, and A. W. W. Ludwig, 
Phys. Rev. B {\bf 78} , 195125 (2008) [arXiv:0803.2786].


\bibitem{shifman-3}
M. Vysotskii, Y. Kogan, and M. Shifman,
Sov. J. Nucl. Phys. {\bf 42} (1985) 318.

\bibitem{peskin}
M. Peskin, Nucl. Phys. {\bf B175} (1980) 197;
S. Dimopoulos, Nucl. Phys. {\bf B168} (1980) 69.

 



\bibitem{V}
  J.~J.~M.~Verbaarschot,
  Phys.\ Rev.\ Lett.\  {\bf 72}, 2531 (1994)
  [arXiv:hep-th/9401059].

\bibitem{VZ3}
 J.~J.~M.~Verbaarschot and I.~Zahed,
  Phys.\ Rev.\ Lett.\  {\bf 73}, 2288 (1994)
  [arXiv:hep-th/9405005].

\bibitem{magnea1} 
U.~Magnea,
  Phys.\ Rev.\  D {\bf 61}, 056005 (2000)
  [arXiv:hep-th/9907096].

\bibitem{magnea4}
U.~Magnea,
  Phys.\ Rev.\  D {\bf 62}, 016005 (2000)
  [arXiv:hep-th/9912207].

\bibitem{dyson-3}
F.J. Dyson, J. Math. Phys. {\bf 3}, 1199 (1962).

\bibitem{class}
  M.~R.~Zirnbauer,
  J.\ Math.\ Phys.\  {\bf 37}, 4986 (1996).
   [arXiv:math-ph/9808012].



\bibitem{VZ}
J.~J.~M.~Verbaarschot and I.~Zahed,
Phys.\ Rev.\ Lett.\  {\bf 70}, 3852 (1993) [arXiv:hep-th/9303012].

\bibitem{az}
  A.~Altland and M.~R.~Zirnbauer,
  Phys.\ Rev.\ B {\bf 55}, 1142 (1997).
[arXiv:cond-mat/9602137].

\bibitem{Mehtabook}
M. L. Mehta, \textit{Random Matrices} (Academic Press Inc., New York, 3rd ed., 2004).

\bibitem{witten2da}
 E.~Witten,
  Commun.\ Math.\ Phys.\  {\bf 141}, 153 (1991).



\bibitem{smilga} 
  A.~V.~Smilga,
  Phys.\ Rev.\ D {\bf 54}, 7757 (1996)
  [arXiv:hep-th/9607007].

\bibitem{smilga-cond} 
  A.~V.~Smilga,
  Phys.\ Rev.\ D {\bf 49}, 6836 (1994)
  [hep-th/9402066].

\bibitem{nagao} 
  T.~Nagao and P.~J.~Forrester,
  Nucl.\ Phys.\ B {\bf 435}, 401 (1995).

\bibitem{ivanov}
 D. Ivanov, J. Math. Phys. {\bf 43}, 126 (2002)
  [arXiv:cond-mat/0103137].
  
\bibitem{akemann-damgaard-jv}
G.~Akemann, D.~Dalmazi, P.~H.~Damgaard and J.~J.~M.~Verbaarschot,
  Nucl.\ Phys.\  B {\bf 601}, 77 (2001)
  [arXiv:hep-th/0011072].

\bibitem{V2}
J.~J.~M.~Verbaarschot,
  Nucl.\ Phys.\  B {\bf 426}, 559 (1994)
  [arXiv:hep-th/9401092].

\bibitem{Gruzberg:2012fb} 
  I.~A.~Gruzberg, A.~D.~Mirlin, and M.~R.~Zirnbauer,
  Phys.\ Rev.\ B {\bf 87}, 125144 (2013)
  [arXiv:1210.6726 [cond-mat.dis-nn]].

\bibitem{supersom}
 H.-J. Sommers,
Act. Phys. Pol. B {\bf 38}, 1001 (2007) [arXiv:0710.5375].

\bibitem{superzirn}
P, Littelmann, H.-J. Sommers, and M.R. Zirnbauer, Math. Phys.
{\bf 283}, 343 (2008) [arXiv:0707.2929].

\bibitem{supermario}
M. Kieburg and T. Guhr, J. Phys. A: Math. Theor. {\bf 42}, 275205 (2009) [arXiv:0905.3253].

\end{thebibliography}
\end{document}